\documentclass[aps,floats,twocolumn,prb,floatfix,showpacs]{revtex4-1}
\usepackage{amssymb}
\usepackage[applemac]{inputenc}
\usepackage{graphicx}
\usepackage{hyperref}
\usepackage{bm}
\usepackage{scrtime}
\usepackage{epstopdf}
\graphicspath{{Figures/}}
\begin{document}

\title{Frustrated local moment models for Fe-pnictide magnetism}

\author{Burkhard Schmidt, Mohammad Siahatgar, and Peter Thalmeier}
\affiliation{Max-Planck-Institut f{\"u}r Chemische Physik fester
Stoffe, 01187 Dresden, Germany}

\date{\today, \thistime}

\begin{abstract}
The low energy spin excitations of the Fe pnictide parent compounds
have been determined by inelastic neutron scattering and interpreted
within the local moment $J_{1a,b}$-$J_2$ Heisenberg model with
orthorhombic symmetry.  This has led to alternative exchange models
that strongly differ in the size of anisotropy.  Although the
compounds are itinerant the localized spin model can explain basic
features of the excitations.  The inherent frustration of this model
leads to quantum fluctuations and possible moment reduction.  We
investigate this question in detail using spin wave approximation and
partly exact diagonalization Lanczos calculations for finite clusters.
We find that the orthorhombic anisotropy stabilizes the columnar AF
phase and its moment.  For the exchange models proposed from inelastic
neutron scattering we can exclude a strong influence of frustration
on the moment size.  We also investigate dependence of magnetization
and susceptibility on field and temperature.
\end{abstract}

\pacs{75.10.Jm, 75.30.Cr, 75.30.Ds}

\maketitle

\section{Introduction}
\label{sect:Introduction}

The discovery of Fe pnictide superconductors has given new impetus to
study the interrelation between magnetic and superconducting
instabilities in condensed matter.  Previous investigation of strongly
correlated heavy fermion and cuprate compounds have also shown this
connection.  However, unlike those compounds the pnictides are only
moderately correlated \cite{yang:09} and in contrast to cuprates
already the parent compounds are metallic.  At lower temperatures they
exhibit a structural phase transition from tetragonal to orthorhombic
symmetry and simultaneously or subsequently show magnetic order
\cite{kaneko:08,yildirim:08}.  For 122 and 1111 compounds the latter
is found to be of the columnar AF type in the FeAs planes
corresponding to wave vector {\bf Q} = ($\pi$,0) which is equal to the
nesting vector of hole and electron Fermi surface pockets obtained
from density functional calculations of the electronic structure.  This
commensurate magnetic structure is stable even for a considerable
range of dopings \cite{yaresko:09,han:09}.  However it was noticed
\cite{yaresko:09,han:09} that the size of the staggered moment depends
strongly on the details of the calculation, especially on the
out-of-plane position of As atoms, and is much larger as the
experimental values obtained from neutron diffraction which is less
than $1\mu_B$ per Fe.  The moments are oriented parallel to the
ordering wave vector which is aligned to the long (a) axis.

On the other hand the results of inelastic neutron scattering (INS) on
the the low energy spin excitations have shown that they may be
successfully described within a localized Heisenberg model that
includes interactions up to next nearest neighbors, i.e., exchange
bonds along the sides and diagonals of the FeAs layers
\cite{ewings:08,mcqueeney:08,zhao:08,diallo:09,zhao:09}.  Therefore
the localized Heisenberg model is of the anisotropic $J_{1a,b}$-J$_2$
type with different exchange constants along the orthorhombic a,b
axes.  The local moment picture has a further interesting aspect:
Depending on the ratios $J_{1a,b}/J_2$ the magnetism may exhibit
frustration which can strongly reduce the size of the ordered moment.
As mentioned above the observed moments are much smaller than the
calculated ones.  Experimentally they vary from $0.36\mu_{\text B}$ in
LaFeAsO to $1\mu_{\text B}$ in SrFe$_2$As$_2$ whereas LDA calculations
give $1.9\mu_{\text B}$ \cite{yin:08} and $1.7\mu_{\text B}$
\cite{han:09} respectively if the experimental values for the As
positions are used.  This translates to a relative moment reduction
factor of 0.19 to 0.59.  Within a model that includes only spin
degrees of freedom one may conjecture that magnetic frustration and
associated enhanced quantum fluctuations are the source of the
strongly reduced ordered moments in the Fe pnictides.  Such an idea,
however needs to be treated with care.  Firstly frustration itself is
not a well defined concept for itinerant spins, but rather refers to
the local moment model.  If the latter is indeed used the extent of
frustration depends crucially on the ratios of exchange constants and
their anisotropies.

No consensus on the proper exchange model in the local moment picture
of Fe pnictides has emerged. Basically two proposals, both from INS
\cite{ewings:08,mcqueeney:08,zhao:08,diallo:09,zhao:09} and local density
calculations \cite{yaresko:09,han:09} have been made: In the first
choice $J_{1b}\simeq J_{1a}$, in the second one 
$|J_{1b}|\ll J_{1a}$. In principle this will lead to very different
dispersion for wave vectors along the $b^*$-axis in the two cases which
may be checked experimentally. But the case remains undecided sofar.
The second very anisotropic case ($J_{1b}$
was even reported slightly ferromagnetic \cite{han:09}) is hard to
understand in terms of a small tetragonal-to orthorhombic structural
distortion with $(a-b)/a\simeq 0.5\cdot 10^{-2} $. This indicates
that the exchange constant should not be interpreted as bond exchange
energies. Rather they are fit parameters obtained from mapping the
total LDA energy with spiral magnetic structure to the classical ground state
energy of a localized Heisenberg model \cite{yaresko:09,han:09}. 

One important motivation why a localized model is nevertheless
worthwhile to study was provided by the INS results.  In a metal one
would naively expect that spin waves with larger wave vectors should
quickly become overdamped when they merge with the continuum of
particle hole excitations.  In fact it was found that well defined
spin waves exist throughout the Brillouin zone \cite{zhao:09} giving
support for the local moment picture.  Even in weakly correlated
$3d$-compounds local moments may be stabilized due to Hund's rule
coupling in a multi-orbital case like Fe pnictides as has been shown
in Ref.~\onlinecite{zhou:09}.  This is not unlike the situation in
elemental Fe ferromagnetism where it was found that Hund's rule
exchange stabilizes the local moments \cite{stollhoff:81,oles:81}.  In
fact spin waves in elemental fcc Fe also exist throughout the
Brillouin zone and may be described by a localized Heisenberg model
although Fe is a good metal \cite{collins:69}.  The itinerant nature
of magnetism in the Fe pnictides does not by itself exclude that a
local moment model is a good starting point for studying the low lying
spin excitations.  Further support for this conjecture was given by
functional renormalization group (FRG) calculations for the isotropic
($J_{1a}=J_{1b}=J_1$) model \cite{zhai:09}.  Starting from a
multiorbital extended Hubbard model it was shown that the $J_1$-$J_2$
model is a valid description for the dominant low energy correlations
for a wide range of parameters.  Coexistence models for both
localized and itinerant moments in the pnictides have been proposed in
Refs.~\onlinecite{kou:09,medici:09}.

Extended models including orbital degrees of freedom within a
localized Kugel-Khomskii type approach have been proposed in Refs.
\onlinecite{krueger:09,*lv:10, chen:09, si:08, rodriguez:09,*rodriguez:10}.  In this case
the ground state can exhibit orbital order which may lead to effective
orthorhombic exchange anisotropy and low ordered moment.  Furthermore
itinerant multi-orbital models \cite{yu:09,eremin:09,lee:09a}, also
including the effect of orbital order
\cite{kubo:09,zhou:09,daghofer:09} for the magnetic ground state have
been proposed.

The INS experiments
\cite{ewings:08,mcqueeney:08,zhao:08,diallo:09,zhao:09} were
interpreted within models including only spin degrees of freedom.
Since we want to refer the exchange parameters found there we also
restrict to these type of models.  In this work we have made a
systematic survey of the anisotropic (orthorhombic) two dimensional
$J_{1a,b}$-$J_2$ Heisenberg model.  Our main subject is to examine
carefully to which extent the `frustration' of nearest and next
nearest exchange constants in this model plays a role in the reduction
of the staggered moment as observed in neutron diffraction.  For that
purpose we are using both analytical spin wave calculations and
numerical Lanczos method for finite clusters of the two-dimensional
(2D) rectangular lattice.  In the central part we investigate the
evolution of the staggered moment, reduced by quantum fluctuations as
function of the frustration and anisotropy ratios.  This allows us to
make a quantitative evaluation of the importance of frustration in the
local moment model for Fe pnictides by comparing with the results for
the experimental exchange constants.  The isotropic model with
$J_{1a,b}=J_1$ has been studied previously within various
approximations \cite{uhrig:09,yao:08} and has also been extended
including interlayer coupling\cite{yao:09,smerald:09}.  Spin waves for
the anisotropic model in zero field are also discussed in
\onlinecite{applegate:10}.

In Sect. \ref{sect:model} we introduce the localized spin model for
the 2D orthorhombic Fe pnictide layers and discuss its
parametrization. In Sect. \ref{sect:phase} we calculate the ground
state energy and phase diagram of the model and the corresponding
location of known Fe pnictide compounds. The reduction of the ordered
moment by quantum fluctuations using spin wave expansion is discussed in detail
 in Sect.~\ref{sect:SW}. The effect of an external field in the anisotropic model is addressed in 
Sect.~\ref{sect:Magnetic}. 
Finally Sect. \ref{sect:Conclusion} gives the summary and conclusion.

\section{Localized spin models for 2D rectangular lattice}

\label{sect:model}
\noindent

We start from a localized spin model with effective spin size $S=1/2$.
The latter is suggested by INS and LDA results given in
Table~\ref{tbl:xchng} which are compatible with this value.  A
stronger argument is given by the Gutzwiller approach to the
multiorbital Hubbard model \cite{zhou:09} which suggests that $2S\simeq
1$ for reasonable model parameters.

The orthorhombic symmetry allows for different n.n. exchange
parameters $J_{1a,b}$, however the extreme difference for some
parameter sets in Table~\ref{tbl:xchng} can hardly be justified by the
simple effect of exchangestriction on nearest-neighbor (n.n.) exchange
caused by the orthorhombic distortion.  As mentioned above, a more
likely source of large exchange anisotropy is the presence of
underlying orbital order which may appear simultaneously with magnetic
order.  However a quantitative prediction of the amount of anisotropy
seems difficult.  Due to the large differences in the proposed
exchange parameters we treat the anisotropy as a free parameter.

INS results also show the existence of a small spin gap of the order
$\simeq 10\,\text{meV}$.  For a $S=1/2$ system this can most easily be
modeled by a uniaxial out-of-plane {\em spin-space\/} exchange
anisotropy which will be included in the model for completeness but
not discussed in detail.  Further insight in the underlying frustrated
exchange model may be gained from field field dependence of
magnetization and susceptibility as has been shown for a different
class of compounds \cite{schmidt:07b,thalmeier:08}.  Therefore we also
include a Zeeman term in the model.

The effective localized spin Hamiltonian we shall discuss in this
paper then has the form
\begin{equation}
    {\cal H}=\sum_{\left\langle ij\right\rangle}
    \vec S_{i}J_{ij}\vec S_{j}
    -g\mu_{\text B}\vec H\sum_{i}\vec S_{i}
    \label{eqn:ham}
\end{equation}
where the sum in the first term extends over bonds $\left\langle
ij\right\rangle$ connecting sites $i$ and $j$. We assume an 
interaction in spin space of the form
\begin{equation}
    J_{ij}=
    \mathop{\rm diag}\nolimits
    \left(J_{ij}^{\perp},J_{ij}^{\perp},J_{ij}^{z}\right).
    \label{eqn:jij}
\end{equation}
To conserve U(1) symmetry, the magnetic field points into the $z$
direction defined by the anisotropy introduced above.  Suppressing the
direction index we set
\begin{equation}
    J_{ij}=\left\{
    \begin{array}{r@{\quad\mbox{if}\quad}l}
	J_{1a}&\vec R_{j}=\vec R_{i}\pm\vec e_{x}
	\\
	J_{1b}&\vec R_{j}=\vec R_{i}\pm\vec e_{y}
	\\
	J_{2}&\vec R_{j}=\vec R_{i}\pm\vec e_{x}\pm\vec e_{y}
	\end{array}
	\right.,
	\label{eqn:js}
\end{equation}
i.\,e., we restrict Eqs.~(\ref{eqn:ham}) and~(\ref{eqn:jij}) to
nearest- and next-nearest neighbor exchange on a rectangular lattice.
For the discussion of the complete phase diagram of this Hamiltonian,
we use a more convenient parameterization of the exchange terms and
write
\begin{eqnarray}
    J_{1a}&=&\sqrt{2}J_{\text c}\cos\phi\cos\theta,
    \nonumber\\
    J_{1b}&=&\sqrt{2}J_{\text c}\cos\phi\sin\theta,
    \\
    J_{2}&=&J_{\text c}\sin\phi,
    \nonumber\\
    J_{\text c}&=&\sqrt{\frac{1}{2}\left(J_{1a}^{2}+J_{1b}^{2}\right)
    +J_{2}^{2}},
    \nonumber
    \label{eqn:exndef}
\end{eqnarray}
introducing an overall energy scale $J_{\text c}$ (this should not be
confused with $J^z$), a frustration angle $\phi$, and an anisotropy
parameter $\theta$.  For $\theta=\pi/4$, the above Hamiltonian reduces
to the square-lattice case $(J_{1a}=J_{1b})$ investigated before, e.g.
in Refs.~\onlinecite{shannon:04,schmidt:07b,thalmeier:08} for V
oxides.

\section{Classical phases and ground state energies}
\label{sect:phase}

For the isotropic model ($\theta=\pi/4$) these are well known
and serve as starting point for discussing frustration effects.  We
first find out how their energies and existence regions are modified
for the general anisotropic case with $-1<\theta/\pi <1$.  On each
site $i$, we introduce a local coordinate system, where the $z'$ axis
is oriented parallel to the local magnetic moment, and we have
\begin{widetext}
\begin{equation}
    \left(
    \begin{array}{c}
	S_{i}^{x}\\
	S_{i}^{y}\\
	S_{i}^{z}
    \end{array}
    \right) =
    \left(
    \begin{array}{ccc}
	\cos(\vec Q\vec R_{i}) & -\sin(\vec Q\vec R_{i}) & 0\\
	\sin(\vec Q\vec R_{i}) & \cos(\vec Q\vec R_{i}) & 0\\
	0 & 0 & 1
    \end{array}
    \right)
    \left(
    \begin{array}{ccc}
	\cos\Theta & 0 & -\sin\Theta\\
	0 & 1 & 0\\
	\sin\Theta & 0 & \cos\Theta
    \end{array}
    \right)
    \left(
    \begin{array}{c}
	S_{i}^{x'}\\
	S_{i}^{y'}\\
	S_{i}^{z'}
    \end{array}
    \right)
    \label{eqn:sprime}
\end{equation}
\end{widetext}
with the ordering vector $\vec Q$ in the $xy$ plane perpendicular to
the magnetic field which points along the $z$ axis.  At finite values
of $\vec H$, the spins form an umbrella-like structure around the
direction of $\vec H$; the respective canting angle $\Theta$ (not to
be confused with the anisotropy parameter $\theta$) is measured
relative to the field direction (global $z$ axis); $\Theta=0$
corresponds to the fully polarized state, and $\Theta=\pi/2$ to the
state with vanishing magnetic field.

With $h=g\mu_{\text B}H$, the classical Hamiltonian then reads
\begin{equation}
    {\cal H}_{\text{cl}}
    =
    NS^{2}
    \left[
    J_{\perp}(\vec Q)
    +A(0)\cos^{2}\Theta
    -\frac{h}{S}\cos\Theta
    \right]
    \label{eqn:hcl}
\end{equation}
where we have introduced the Fourier transform
\begin{equation}
    J_{\alpha}(\vec k)=\frac{1}{N}
    \sum_{\langle ij\rangle}J_{ij}^{\alpha}
    e^{-{\rm i}\vec k(\vec R_{i}-\vec R_{j})}
    =\frac{1}{2}\sum_{n}J_{n}^{\alpha}e^{-{\rm i}\vec k\vec R_{n}}
    \label{eqn:jkdef}
\end{equation}
for $\alpha=\{\perp,z\}$ and the last sum runs over all bonds $n$
connecting a fixed site $i$ with its neighbors. The coefficient 
$A(0)=A(\vec k=0)$ is defined by
\begin{equation}
    A(\vec k)=J_{z}(\vec k)
    +\frac{1}{2}
    \left(
    J_{\perp}(\vec k+\vec Q)+J_{\perp}(\vec k-\vec Q)
    \right)
    -2J_{\perp}(\vec Q).
    \label{eqn:ak}
\end{equation}
The reason for defining $A(\vec k)$ in this way will become clear
later in Sec.~\ref{sect:SW}.  Minimizing Eq.~(\ref{eqn:hcl}) with
respect to $\Theta$, we get the classical canting angle $\Theta_{\text
c}$ via
\begin{equation}
    \cos\Theta_{\text c}=\frac{h}{2SA(0)},
    \label{eqn:thetaca0}
\end{equation}
and Eq.~(\ref{eqn:hcl}) reads
\begin{equation}
    {\cal H}_{\text{cl}}
    =
    NS^{2}\left[
    J_{\perp}(\vec Q)
    -A(0)\cos^{2}\Theta_{\text c}
    \right].
    \label{eqn:hcl0}
\end{equation}
Minimizing this Hamiltonian with the exchange parameters from 
Eq.~(\ref{eqn:js}) with respect to the components of $\vec Q$
leads to the four well-known classical phases with ordering vectors
\begin{equation}
    \vec Q=\left\{
    \begin{array}{l@{\quad}l}
	0&\mbox{ferromagnet (FM)}\\
	(\pi/a,\pi/b)&\mbox{Néel antiferromagnet (NAF)}\\
	(\pi/a,0)&\mbox{columnar AF along $a$ (CAFa)}\\
	(0,\pi/b)&\mbox{columnar AF along $b$ (CAFb)}
    \end{array}
    \right..
\end{equation}
The minimization condition reduces to $\partial J_{\perp}(\vec
Q)/\partial\vec Q=0$ and is thus field independent and depends on the
transverse exchange parameters only.  The classical ground-state energies are
\begin{widetext}
\begin{equation}
    E_{\text{gs}}^{\text{cl}}=NS^2\left\{
      \begin{array}{l@{\qquad}l}
	  J_{1a}^{\perp}+J_{1b}^{\perp}+2J_{2}^{\perp}&\mbox{FM}\\
	  2J_{2}^{\perp}
	  -\left(
	  J_{1a}^{\perp}+J_{1b}^{\perp}
	  \right)
	  +
	  \left[
	  2\left(J_{2}^{\perp}-J_{2}^{z}\right)
	  -\left(J_{1a}^{\perp}+J_{1a}^{z}\right)
	  -\left(J_{1b}^{\perp}+J_{1b}^{z}\right)
	  \right]
	  \cos^{2}\Theta_{\text c}
	  &\mbox{NAF}\\
	  J_{1b}^{\perp}
	  -\left(
	  J_{1a}^{\perp}+2J_{2}^{\perp}
	  \right)
	  +
	  \left[
	  \left(J_{1b}^{\perp}-J_{1b}^{z}\right)
	  -\left(J_{1a}^{\perp}+J_{1a}^{z}\right)
	  -2\left(J_{2}^{\perp}+J_{2}^{z}\right)
	  \right]
	  \cos^{2}\Theta_{\text c}
	  &\mbox{CAFa}\\
	  J_{1a}^{\perp}
	  -\left(
	  J_{1b}^{\perp}+2J_{2}^{\perp}
	  \right)
	  +
	  \left[
	  \left(J_{1a}^{\perp}-J_{1a}^{z}\right)
	  -\left(J_{1b}^{\perp}+J_{1b}^{z}\right)
	  -2\left(J_{2}^{\perp}+J_{2}^{z}\right)
	  \right]
	  \cos^{2}\Theta_{\text c}
	  &\mbox{CAFb}
      \end{array}
    \right.
    \label{eqn:egscl}
\end{equation}
where $\cos\Theta_{\text c}=h/h_{\text s}$, and the critical or saturation fields 
for the nonuniform phases are given by Eq.~(\ref{eqn:thetaca0}),
\begin{equation}
    \frac{h_{\text s}}{2S}=
    J_{1a}^{z}+J_{1b}^{z}+2J_{2}^{z}
    -\left\{
      \begin{array}{l@{\qquad}l}
	  2J_{2}^{\perp}-\left(J_{1a}^{\perp}+J_{1b}^{\perp}\right)&\mbox{NAF}\\
	  J_{1b}^{\perp}-\left(J_{1a}^{\perp}+2J_{2}^{\perp}\right)&\mbox{CAFa}\\
	  J_{1a}^{\perp}-\left(J_{1b}^{\perp}+2J_{2}^{\perp}\right)&\mbox{CAFb}
      \end{array}
    \right..
\end{equation}
\end{widetext}

The minimization condition contains an additional extremal solution having an 
incommensurate wave vector given by
\begin{equation}
    \cos Q_{x}a = -\frac{J_{1b}}{2J_{2}},\quad
    \cos Q_{y}b = -\frac{J_{1a}}{2J_{2}},
\end{equation}
with a ground state energy
\begin{equation}
    E_{\text{gs}}^{\text{cl}} = -\frac{J_{1a}J_{1b}}{8J_{2}}.
\end{equation}
However, this energy for the incommensurate wave vector is always
higher than or equal to the energy in Eq.~(\ref{eqn:egscl}) of the
commensurate ground state corresponding to the values chosen for the
exchange constants.

From the classical ground state energy in Eq.~(\ref{eqn:egscl}) one may 
already construct the phase diagram in the $\phi,\theta$ plane, however 
for the following discussions it is important to include the effect of 
quantum fluctuations.

\section{Quantum fluctuations and ordered moment size in spin wave
approximation}
\label{sect:SW}

In the regions of the $\phi,\theta$ plane where two or more of the
classical phases become degenerate large quantum fluctuations appear
and reduce or suppress the ordered moment.  These are the strongly
frustrated regions of the phase diagram.  One may approach them to
some extent by starting from the stable region, and calculate the
contribution of zero point fluctuations in spin wave approximation.
This leads to an improved ground state energy, an estimate for the
reduction of the ordered moment and for the extent of the instability
region where magnetic order breaks down.  This program has been
successfully implemented before for the isotropic case
\cite{schmidt:07b,thalmeier:08}.  It will now be carried out for the
more general model in order to quantify the importance of frustration
and quantum fluctuations for the compounds listed in Table
\ref{tbl:xchng}.

Returning to Eq.~(\ref{eqn:ham}) expressed in the local coordinate
system introduced in Sec.~\ref{sect:phase}, Eq.~(\ref{eqn:sprime}), we
apply a Holstein-Primakoff transformation and carry out a $1/S$
expansion, keeping terms up to first order in $1/S$.  (We regard $h$
formally as proportional to $S$.)  Next, we apply a Fourier
transformation.  A detailed description of the necessary steps can be
found in Appendix~\ref{app:hp}.  The resulting Hamiltonian is given by
\begin{widetext}
\begin{eqnarray}
   {\cal H}&=&
   {\cal H}_{\text{cl}}+
   NS
   \left[
   J_{\perp}(\vec Q)
   +A(0)\cos^{2}\Theta
   -\frac{h}{2S}\cos\Theta
   \right]
   +\sqrt{2NS^{3}}\sin\Theta
   \left[A(0)\cos\Theta-\frac{h}{2S}\right]
   \left(a_{0}+a_{0}^{\dagger}\right)
   \nonumber\\&&
   +\frac{S}{2}\sum_{\vec k}
   \left\{
   \left[
   A(\vec k)
   -\cos^{2}\Theta\left(B(\vec k)+2A(0)\right)
   +\frac{h}{S}\cos\Theta
   \right]
   \left(
   a_{\vec k}^{\dagger}a_{\vec k}
   +a_{-\vec k}a_{-\vec k}^{\dagger}
   \right)
   \right.
   \nonumber\\&&
   \phantom{+\frac{S}{2}\sum_{\vec k}}
   \left.
   +B(\vec k)\left(1-\cos^{2}\Theta\right)
   \left(
   a_{\vec k}a_{-\vec k}+a_{\vec k}^{\dagger}a_{-\vec k}^{\dagger}
   \right)
   +C(\vec k)\cos\Theta 
   \left(
   a_{\vec k}^{\dagger}a_{\vec k}
   -a_{-\vec k}a_{-\vec k}^{\dagger}
   \right)    
   \right\}
   \label{eqn:hlswb}
\end{eqnarray}
\end{widetext}
where $A(\vec k)$ is defined in Eq.~(\ref{eqn:ak}), and
\begin{eqnarray}
    B(\vec k)&=&J_{z}(\vec k)
    -\frac{1}{2}
    \left(
    J_{\perp}(\vec k+\vec Q)+J_{\perp}(\vec k-\vec Q)\right),
    \label{eqn:bk}
    \\
    C(\vec k)&=&J_{\perp}(\vec k+\vec Q)-J_{\perp}(\vec k-\vec Q).
    \label{eqn:ck}
\end{eqnarray}
Eq.~(\ref{eqn:hlswb}) still contains a part which is linear in the
bosons with zero momentum.  It occurs only in finite magnetic fields.
In equilibrium, when $\Theta=\Theta_{\text c}$ (see
Eq.~(\ref{eqn:thetaca0})), this part vanishes, and we will drop it in
the following.  A subsequent Bogoliubov transformation leads to the
final form
\begin{equation}
    {\cal H}={\cal H}_{\text{cl}}+{\cal H}_{\text{zp}}
    +S\sum_{\vec k}E(h,\vec k)\alpha_{\vec k}^{\dagger}\alpha_{\vec 
    k},
    \label{eqn:hbog}
\end{equation}
where ${\cal H}_{\text{cl}}$ is given by Eq.~(\ref{eqn:hcl0}), and 
\begin{equation}
    {\cal H}_{\text{zp}}=
    NS
    J_{\perp}(\vec Q)
    +\frac{S}{2}\sum_{\vec k}E(h,\vec k)
    \label{eqn:hzp}
\end{equation}
is the zero-point energy contribution to the total ground state energy
$E_{\text{gs}}={\cal H}_{\text{cl}}+{\cal H}_{\text{zp}}$.  The former
corresponds to the magnon excitations described by the boson operators
\begin{eqnarray}
    \alpha_{\vec k}&=&
    u_{\vec k}a_{\vec k}+v_{\vec k}a_{-\vec k}^{\dagger},
    \label{eqn:alpha}
    \\
    \alpha_{-\vec k}^{\dagger}&=&
    v_{\vec k}a_{\vec k}+u_{\vec k}a_{-\vec k}^{\dagger}.
    \label{eqn:alphad}
\end{eqnarray}
The $\vec k$ sums in the equations above span the full
crystallographic Brillouin zone.  For the spin-wave dispersion, we
obtain the expression
\begin{widetext}
\begin{equation}
    E(h,\vec k)=
    \sqrt{
    \left[
    A(\vec k)-B(\vec k)\cos^{2}\Theta_{\text c}
    \right]^{2}
    -\left[
    B(\vec k)\left(1-\cos^{2}\Theta_{\text c}\right)
    \right]^{2}
    }
    +C(\vec k)\cos\Theta_{\text c}.
    \label{eqn:ek}
\end{equation}
\end{widetext}
$C(\vec k)$ only occurs at finite magnetic fields, and because it is
antisymmetric in $\vec k$, it does not contribute to the zero-point
fluctuations.

\subsection{Total ground-state energy}

We now calculate the total ground state energy in spin wave approximation
to determine the zero-field phase diagram. We also will give a comparison to the 
classical ground state energy and the results for finite clusters obtained
from the exact diagonalization Lanczos method. 

Unless explicitly mentioned otherwise, we assume $\Theta=\Theta_{\text
c}$ from here on.  Furthermore spin space anisotropy is ignored
($J_{ij}^{\perp}=J_{ij}^{z}$) from now on.  The ground-state energy is
given by the sum of the classical energy, Eq.~(\ref{eqn:hcl0}) and the
zero-point fluctuations of the magnons, Eq.~(\ref{eqn:hzp}), with the
dispersion from Eq.~(\ref{eqn:ek}).  Explicitly we have for isotropic
exchange parameters
\begin{eqnarray}
    \lefteqn{\mbox{FM: no zero-point fluctuations,}}\hskip2em
    \nonumber\\
    J(\vec Q)&=&J_{1a}+J_{1b}+2J_{2},
    \nonumber\\
    A(\vec k)&=&
    2\left[
    J_{1a}\left(\cos(k_{x}a)-1\right)
    +J_{1b}\left(\cos(k_{y}b)-1\right)
    \right.
    \nonumber\\&&{}
    \left.
    +2J_{2}\left(\cos(k_{x}a)\cos(k_{y}b)-1\right)
    \right],
    \label{eqn:efm}\\
    B(\vec k)&=&0,
    \nonumber\\[\baselineskip]
    \lefteqn{\mbox{NAF:}\ \vec Q=(\pi,\pi),}\hskip2em
    \nonumber\\
    J(\vec Q)&=&-\left(J_{1a}+J_{1b}\right)+2J_{2},
    \nonumber\\
    A(\vec k)&=&
    2\left[
    J_{1a}
    +J_{1b}
    +2J_{2}\left(\cos(k_{x}a)\cos(k_{y}b)-1\right)
    \right],
    \nonumber\\
    B(\vec k)&=&
    -2\left[J_{1a}\cos(k_{x}a)+J_{1b}\cos(k_{y}b)\right],
    \label{eqn:enaf}\\[\baselineskip]
    \lefteqn{\mbox{CAFa:}\ \vec Q=(\pi,0),}\hskip2em
    \nonumber\\
    J(\vec Q)&=&J_{1b}-\left(J_{1a}+2J_{2}\right),
    \nonumber\\
    A(\vec k)&=&
    2\left[
    J_{1a}
    +J_{1b}\left(\cos(k_{y}b)-1\right)
    +2J_{2}
    \right],
    \label{eqn:ecafb}\\
    B(\vec k)&=&
    -2\left[J_{1a}\cos(k_{x}a)+2J_{2}\cos(k_{x}a)\cos(k_{y}b)\right].
    \nonumber\\[\baselineskip]
    \lefteqn{\mbox{CAFb:}\ \vec Q=(0,\pi),}\hskip2em
    \nonumber\\
    J(\vec Q)&=&J_{1a}-\left(J_{1b}+2J_{2}\right),
    \nonumber\\
    A(\vec k)&=&
    2\left[
    J_{1a}\left(\cos(k_{x}a)-1\right)
    +J_{1b}
    +2J_{2}
    \right],
    \label{eqn:ecafa}\\
    B(\vec k)&=&
    -2\left[J_{1b}\cos(k_{y}b)+2J_{2}\cos(k_{x}a)\cos(k_{y}b)\right],
    \nonumber
\end{eqnarray}

\begin{figure}
    \centering
    \includegraphics[width=.9\columnwidth]{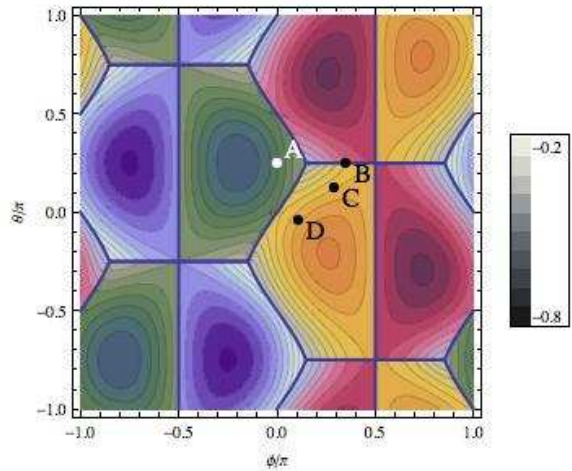}
    \caption{Ground-state energy in linear spinwave approximation of
    the frustrated Heisenberg Hamiltonian on the rectangular lattice
    as a function of the frustration angle $\phi$ and the anisotropy
    parameter $\theta$.  Energy unit is the overall energy scale
    $J_{\text c}$, the magnetic field is zero.  The four different
    classical phases are labeled by color: Blue -- FM, green -- NAF,
    orange -- CAFa, red -- CAFb.  The thick lines correspond to the
    classical phase boundaries, the symbols indicate the locations of
    the parameters used in Fig.~\protect\ref{fig:swspectra} and are
    labeled with the corresponding character, also used in
    Table~\ref{tbl:xchng}.  The white dot represents the standard
    nearest-neighbor Heisenberg model ($J_{1a}=J_{1b}=J_{1}$,
    $J_{2}=0$), the black dots denote experimental points for
    BaFe$_{2}$As$_{2}$ in Ref.~\onlinecite{ewings:08} and
    CaFe$_{2}$As$_{2}$ in Refs.~(\onlinecite{diallo:09,zhao:09}).}
    \label{fig:egslsw}
\end{figure}
Figure~\ref{fig:egslsw} shows a contour plot of the total ground-state
energy $E_{\text{gs}}={\cal H}_{\text{cl}}+{\cal H}_{\text{zp}}$ at
zero field as a function of the frustration angle $\phi$ and the
anisotropy parameter $\theta$.  The energy unit is $J_{\text c}$.
Four magnetic phases appear (see caption) in a symmetric pattern in
the $\phi,\theta$-plane.  We present the complete phase diagram of the
$J_{1a,b}$-$J_2$ model although only the sector $0<\theta/\pi<0.25$,
$0<\phi/\pi<0.5$ seems to be relevant for the Fe pnictide class
according to Table~\ref{tbl:xchng}.  We notice the following
characteristics of the phase diagram:

i) The ground state energy and phase diagram are invariant under the
following symmetry transformations: Reflections at the lines
$\theta=\pi/4$ and $-3\pi/4$ and inversion at the points
$(\phi,\theta)=(\pm\pi/2,3\pi/4)$.  Both operations lead to
$(J_{1a},J_{1b})\rightarrow (J_{1b},J_{1a})$ with $J_2$ unchanged.
This corresponds to an interchange of the the columnar CAFa/b phases
while FM and NAF are mapped identically.  The classical ground state
energy has even more symmetries. 

ii) In the isotropic case ($\theta=\pi/4$) CAFa and CAFb are
degenerate, and moving away from this symmetry line one of the two
phases is selected.  The stability region of the columnar phases along
the frustration axis ($\phi$) increases upon going away from the
symmetry line $\theta=\pi/4$ while that of the neighboring NAF or FM
phase decreases.  Therefore CAFa/b is stabilized by the presence of a
$J_{1a,b}$ anisotropy.

iii) The exchange frustration is largest where three phases (e.g. the
corner point $\theta/\pi=0.25$, $\phi/\pi\simeq 0.15$ and
$J_{2}/J_{1}=1/2$) or two phases (the boundary lines) meet.  Therefore
the degree of frustration in a given compound in Table~\ref{tbl:xchng}
also depends on the size of its anisotropy.  While the above corner
point is strongly frustrated and in fact has no long range order (see
Sect.~\ref{sect:ms}) the point $\theta=0$, $\phi/\pi\approx0.15$
($J_{1b}=0$, close to (D) in Fig.~\ref{fig:egslsw}) is {\em not\/}
strongly frustrated but stable CAFa despite having
$\phi/\pi\approx0.15$ or $J_{2}/(J_{1a}/\sqrt{2})=1/2$.

\begin{figure}
    \centering
    \includegraphics[width=.4\textwidth]{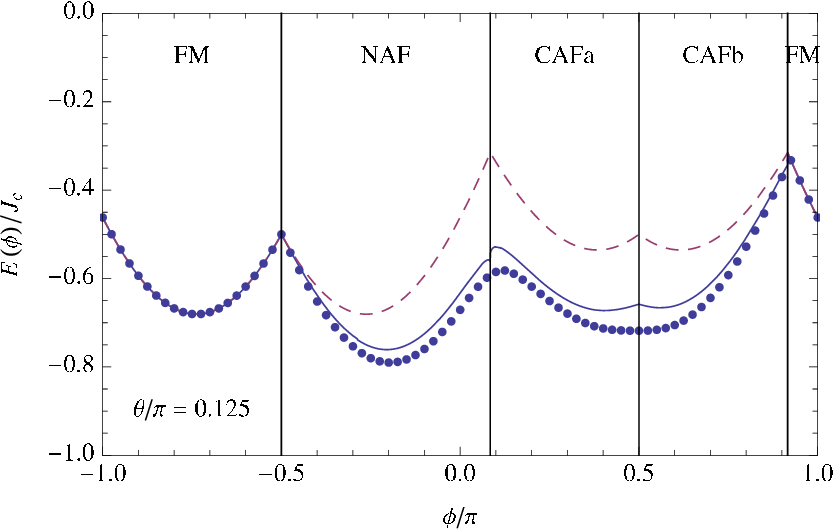}
    \caption{Ground-state energy as a function of the frustration
    angle $\phi$ for fixed anisotropy parameter $\theta=\pi/8$ from
    linear spin-wave theory (solid line) and exact diagonalization
    (solid dots; 24 sites).  For comparison, the dashed curve shows the
    classical ground-state energy.  Here and in subsequent plots of
    $\phi$ dependent quantities the vertical lines indicate the
    classical phase boundaries.}
    \label{fig:egscut}
\end{figure}

Apart from the ferromagnet, which is an eigenstate to the full
Hamiltonian, the spin-wave corrections stabilize the classical ground
state, i.\,e., the zero-point energy in Eq.~(\ref{eqn:hzp}) is
negative semidefinite for all values of $\phi$ and $\theta$.  As an
example, Fig.~\ref{fig:egscut} displays the dependence of the
ground-state energy on the frustration angle $\phi$ obtained both from
linear spin-wave theory (solid line) and from the classical model
(dashed line).  The plot was made with fixed anisotropy parameter
$\theta=\pi/8$, corresponding to a spatial anisotropy
$J_{1b}/J_{1a}=\sqrt{2}-1\approx0.414$.  For a comparison we also
present the numerical results from Lanczos calculations for finite
clusters of the square lattice with size $N=24$ which is quite close
to the spin wave results.

\subsection{Spin-wave spectra}

\begin{table} 
    \centering
    \begin{tabular}{clclccccrr}
	&System&Ref.
	&$S$&$SJ_{1a}$&$SJ_{1b}$&$SJ_2$&$SJ_{\text c}$&$\phi/\pi$&$\theta/\pi$\\ 
	\hline
	1&CaFe$_2$As$_2$ & \onlinecite{mcqueeney:08}
	&--& 41    & 10   & 21    & 36   & 0.19 & 0.08 \\
	C&CaFe$_2$As$_2$ & \onlinecite{diallo:09}
	& 0.4& 24--37   & 7--20   & 28--34   & 33--45   & 0.29 & 0.13 \\ 
	D&CaFe$_2$As$_2$ & \onlinecite{zhao:09}
	& 0.22 & 49.9   & -5.7   & 18.9   & 53.7   & 0.11 & -0.04 \\
	B&BaFe$_2$As$_2$ & \onlinecite{ewings:08}
	&0.28    & 17.5    & 17.5  & 35    & 39.1 & 0.35 & 0.25 \\
	5&BaFe$_2$As$_2$ & \onlinecite{ewings:08}
	&0.54   & 36    & -7   & 18    & 31.6 & 0.19 & -0.06 \\	
	\hline
	6&CaFe$_2$As$_2$ & \onlinecite{han:09}
	&0.75  & 27.4   & -2.1   & 14.5   & 24.3   & 0.20 & -0.02 \\
	7&BaFe$_2$As$_2$ & \onlinecite{han:09}
	&0.84   & 36.1   & -2.6  & 12.0    & 38.0 & 0.10 & -0.02 \\
	8&SrFe$_2$As$_2$ & \onlinecite{han:09}
	&0.84  &  35.3    & 2.2  & 13.4    & 28.4 & 0.16 & 0.02 \\
    \end{tabular} 
    \caption{Fe pnictide moment $\mu=2S\mu_{\text B}$ and exchange
    interactions (in meV) from experiment (top) and theory (bottom).
    Here $J_{\text c}$ is the average exchange energy scale and
    $\theta$, $\phi$ are anisotropy and frustration angles
    (Eq.~(\ref{eqn:exndef})). The first column holds the labels used 
    in Fig.~\ref{fig:egslsw} (letters) and Fig.~\ref{fig:msexp} 
    (letters and numbers).}
    \label{tbl:xchng}
\end{table} 
We shall now discuss the excitation spectrum of the
Hamiltonian~(\ref{eqn:hbog}) for  some typical points in the
$(\phi,\theta)$ phase diagram, Fig.~\ref{fig:egslsw}.  In
Table~\ref{tbl:xchng}, we have compiled an excerpt from the available
literature on experimental and theoretical values for the exchange
parameters of {\em A\/}Fe$_2$As$_2$ compounds, where {\em A\/} denotes
an alkaline metal.  From this list, we have chosen three parameter
sets, indicated by the black dots in Fig.~\ref{fig:egslsw}, and the
standard Néel antiferromagnet (white dot).

\begin{figure*}
    \centering
    \includegraphics[width=.4\textwidth]{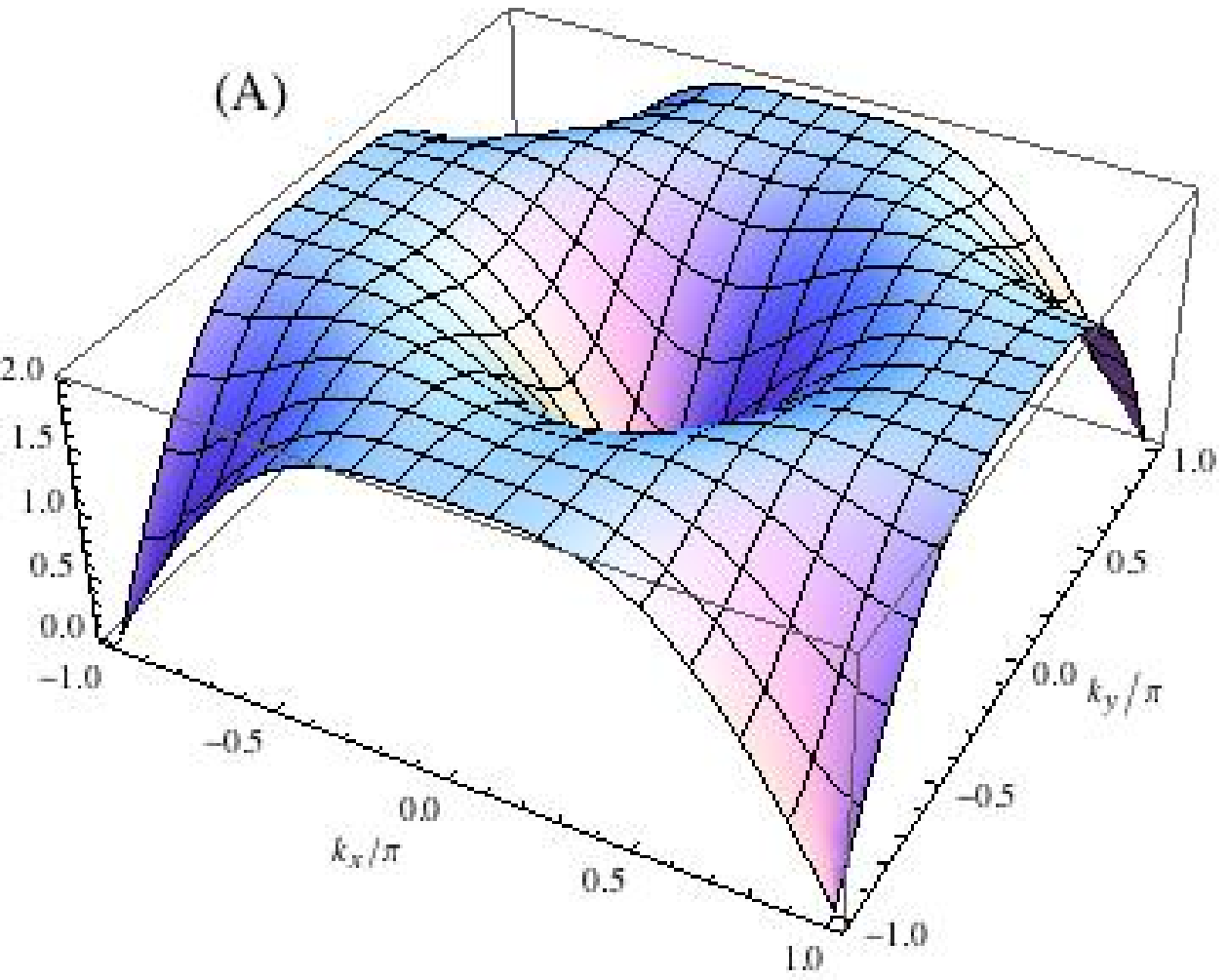}
    \includegraphics[width=.4\textwidth]{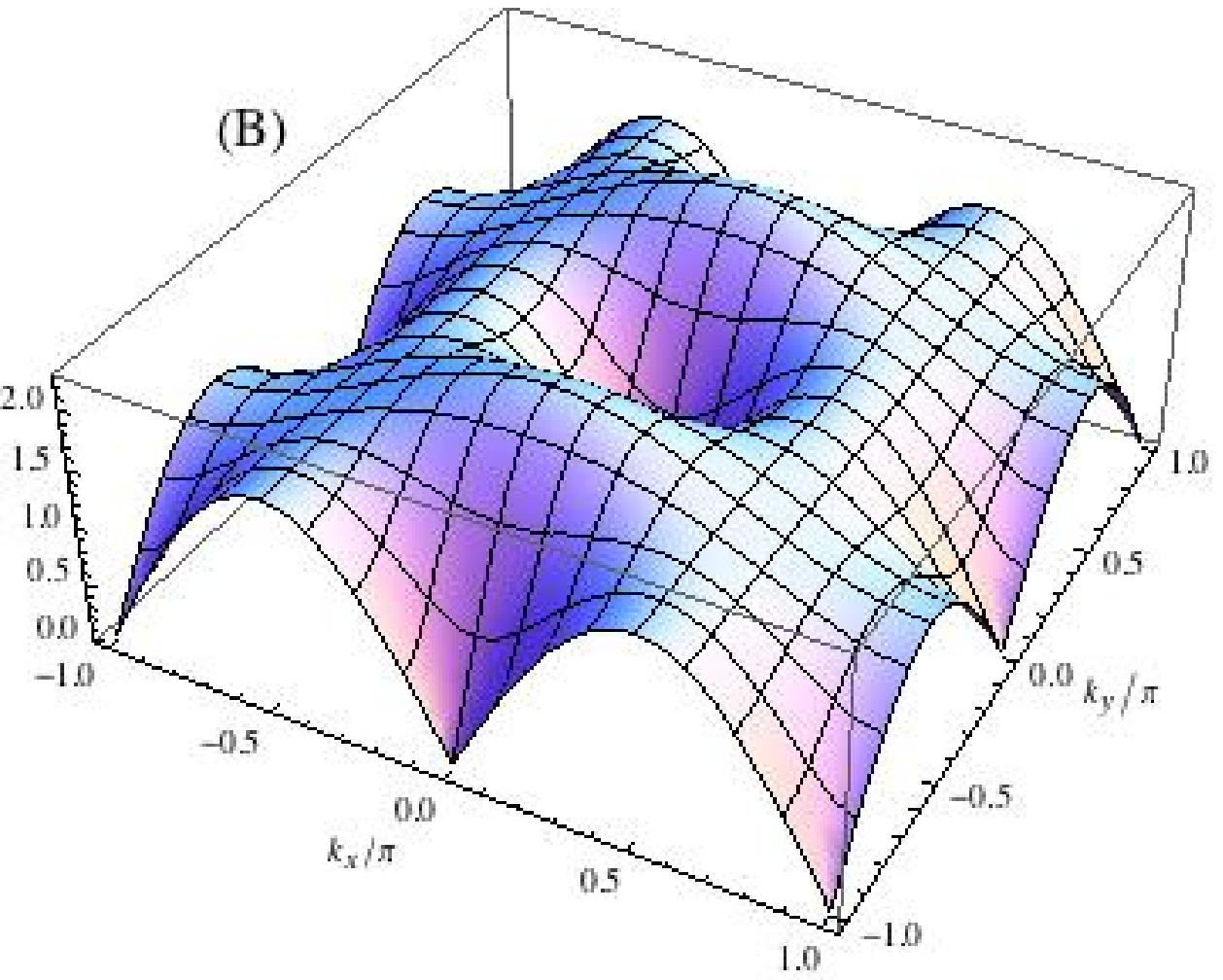}\\
    \includegraphics[width=.4\textwidth]{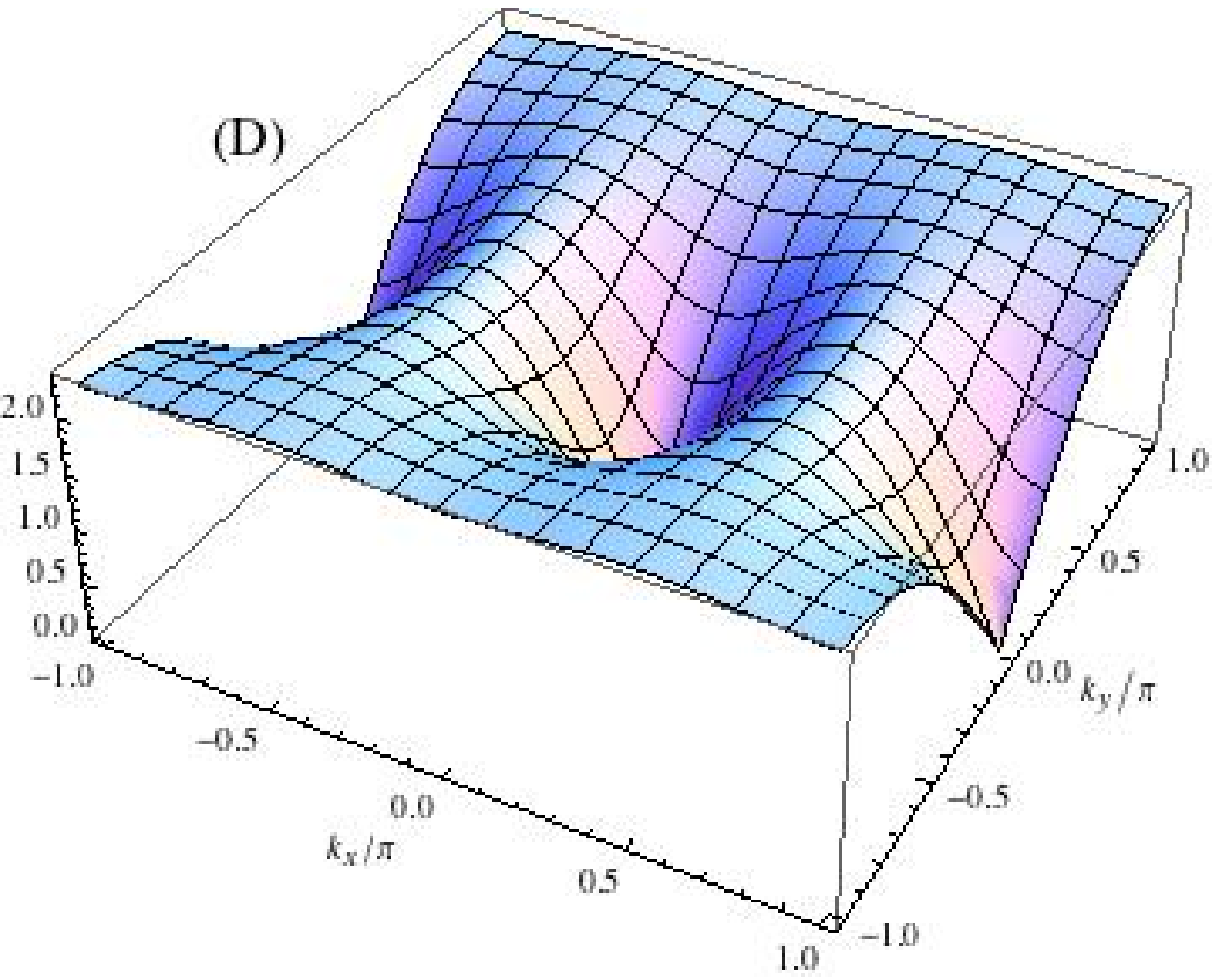}
    \includegraphics[width=.4\textwidth]{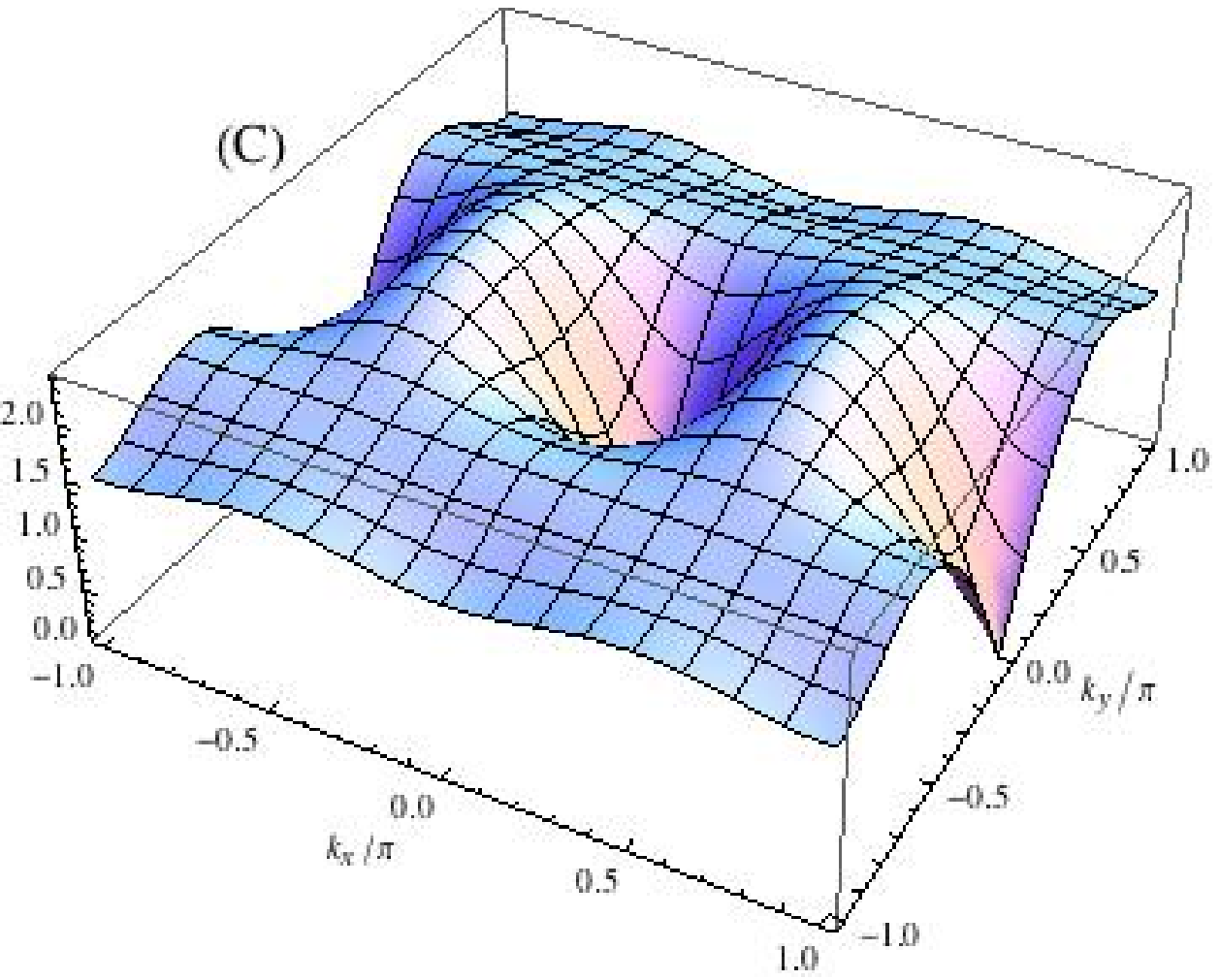}
    \caption{Spin wave spectra $\omega(\vec k)/J_{\text c}$.  Clockwise
    from top left: (A) $(\phi,\theta)/\pi=(0,0.25)$ -- NAF, isotropic
    exchange with $J_{2}=0$; (B) $(\phi,\theta)/\pi=(0.35,0.25)$ --
    CAFa/b, isotropic exchange; (C) $(\phi,\theta)/\pi=(0.29,0.13)$ --
    inside CAFa; (D) $(\phi,\theta)/\pi=(0.11,-0.04)$ -- even more
    inside CAFa.}
    \label{fig:swspectra}
\end{figure*}

Figure~\ref{fig:swspectra} shows plots of the $\vec k$ dependence of
the spin wave excitations for different parameter sets
$(\phi,\theta)$.  The parameter sets used for the plots are indicated
by the symbols in Fig.~\ref{fig:egslsw}.  For simplicity, we scale
$\vec k$ with the lattice constants and set $k_{x}a\to k_{x}$ and
$k_{y}b\to k_{y}$.  All plots refer to the full crystallographic
Brillouin zone.

The top left spectrum (A) in Fig.~\ref{fig:swspectra} shows the
well-known dispersion for the nearest-neighbor Heisenberg model for
comparison.  It has a Goldstone mode at the equivalent wave vectors
$\vec Q=0$ and $\vec Q=(\pm\pi,\pm\pi)$.  The low-energy dispersion $\omega(\vec k)= SE(\vec k)$ is
linear around these points with
\begin{eqnarray}
    \lefteqn{
    \omega(\vec k)=2S
    \sqrt{J_{1a}+J_{1b}}
    }
    \\
    &&
    \times
    \sqrt{
    \left(J_{1a}-2J_{2}\right)
    \left(k_{x}-Q_{x}\right)^{2}
    +
    \left(J_{1b}-2J_{2}\right)
    \left(k_{y}-Q_{y}\right)^{2}
    }
    .
    \nonumber
\end{eqnarray}

The top-right plot (B) shows the dispersion for
$(\phi,\theta)/\pi=(0.35,0.25)$, corresponding to an isotropic
exchange on the border between CAFa and CAFb phases.  These parameters
have been determined for BaFe$_{2}$As$_{2}$ in
Ref.~\onlinecite{ewings:08}.  The Goldstone modes are at $\vec
Q=(0,\pm\pi)$ and $\vec Q=(\pm\pi,0)$ and the equivalent points $\vec
Q=0$ and $\vec Q=(\pm\pi,\pm\pi)$, reflecting the twofold degeneracy
of the CAFa and CAFb phases.  The linear dispersion around the minima
is
\begin{eqnarray}
    \lefteqn{
    \omega_{a}(\vec k)=2S
    \sqrt{2J_{2}+J_{1a}}
    }
    \\
    &&
    \times
    \sqrt{
    \left(2J_{2}+J_{1a}\right)
    \left(k_{x}-Q_{x}\right)^{2}
    +
    \left(2J_{2}-J_{1b}\right)
    \left(k_{y}-Q_{y}\right)^{2}
    }
    \nonumber
\end{eqnarray}
for the CAFa phase, and
\begin{eqnarray}
    \lefteqn{
    \omega_{b}(\vec k)=2S
    \sqrt{2J_{2}+J_{1b}}
    }
    \\
    &&
    \times
    \sqrt{
    \left(2J_{2}-J_{1a}\right)
    \left(k_{x}-Q_{x}\right)^{2}
    +
    \left(2J_{2}+J_{1b}\right)
    \left(k_{y}-Q_{y}\right)^{2}
    }
    \nonumber
\end{eqnarray}
for the CAFb phase.

For $(\phi,\theta)/\pi=(0.29,0.13)$, assigned to CaFe$_{2}$As$_{2}$ in
Ref.~\onlinecite{diallo:09}, we show the spin-wave dispersion in the
bottom right plot (C) of Fig.~\ref{fig:swspectra}.  With these
parameters, the system is deep inside the CAFa phase.  In contrast to
the isotropic case, the dispersion around $\vec Q^{*}=(0,\pm\pi)$ and
$(\pm\pi,\pm\pi)$, while still being local minima (but with a
quadratic $\vec k$ dependence), have a finite energy gap.  We have
$E_{a}(\vec k)=0$ remaining only at the wave vectors $\vec k=0$
and $(\pm\pi,0)$, characteristic for the CAFa phase.

Finally, the bottom left plot (D) in Fig.~\ref{fig:swspectra} displays
the dispersion for $(\phi,\theta)/\pi=(0.11,-0.04)$.  This alternative
parameter set was proposed in Ref.~\onlinecite{zhao:09} for
CaFe$_{2}$As$_{2}$.  The local minima at $\vec Q^{*}=(0,\pm\pi)$ and
$(\pm\pi,\pm\pi)$ discussed in the previous paragraph have almost
disappeared, the dispersion at the zone boundary $k_{y}=\pm\pi$ is
flat.  This property can be utilized to decide between the two
apparently different parameter sets for the identical compound.  In
fact from this comparison it was concluded~\cite{zhao:09} that the
strongly anisotropic set (D) describes the dispersion along $(0,k_y)$
much better for large wavevectors $k_y/\pi>0.5$.  Since the ordering
is still of the CAFa type, the Goldstone mode at $\vec Q=(\pm\pi,0)$
remains for the 2D model. A realistic description of the dispersion requires,
however, the inclusion of interplane exchange \cite{diallo:09,zhao:09} which
leads to a finite gap at these points.\cite{smerald:09}

\subsection{Ordered moment}
\label{sect:ms}

The most appropriate quantity for judging the degree of frustration in
the local moment model is the size of the ordered ground state moment
$m_{\text s}(\phi,\theta)$ relative to its size for the unfrustrated
($J_2=0$) isotropic ($J_{1a}=J_{1b}=J_1$) NAF state.  The latter is
already reduced with respect to the classical value S to $m_{\text
s}^0\sim0.607S$.  The stronger the frustration the more $m_{\text s}$
should be reduced even below the NAF value.  In the isotropic case
when $J_2<0$ there is obviously no frustration and the NAF state is
even stabilized.  The essential question is: how large is the degree
of frustration in the CAFb,b states relevant for the Fe pnictides?
This question can be answered by calculating the moment reduction
$m_{\text s}(\phi,\theta)/S$ in spin wave approximation.

The ordered moment is the ground-state expectation value of the $z$
component of the spin $\vec S'$ in the local coordinate system,
\begin{eqnarray}
    m_{\text s}
    &=&
    \frac{1}{N}\sum_{i}
    \left\langle S_{i}^{z'}\right\rangle
    =
    S-\frac{1}{N}\sum_{\vec k}
    \left\langle\alpha_{\vec k}^{\dagger}\alpha_{\vec k}\right\rangle
    \nonumber\\
    &=&
    S-\frac{1}{N}\sum_{\vec k}v_{\vec k}^{2}.
    \label{eqn:msgen}
\end{eqnarray}
Inserting the expression for $v_{\vec k}$ required to bring $\cal H$ 
into diagonal form (see Appendix~\ref{app:ms} for a complete 
expression), we get
\begin{equation}
    m_{\text s}=S\left[1-\frac{1}{2S}\left(
    \frac{1}{N}\sum_{\vec k}
    \frac{A(\vec k)-B(\vec k)\cos^{2}\Theta_{\text c}}{E(h,\vec k)}
    -1\right)\right].
    \label{eqn:ms}
\end{equation}
Due to quantum fluctuations $m_{\text s}<S$ is smaller than in the
classical case, except for the ferromagnet.  Near the borders of the
CAFb and CAFa phases to the NAF phase, the ordered moment vanishes,
indicating the failure of spin wave theory due to strong frustrations.
Also between the FM and CAF phases the latter lead to a vanishing
$m_{\text s}$ in a small region.

\begin{figure}
    \centering
    \includegraphics[width=0.4\textwidth]{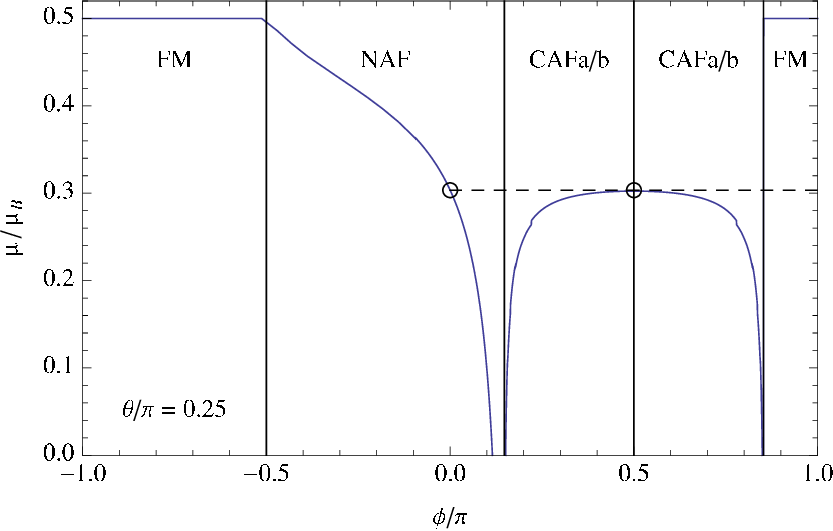}
    \includegraphics[width=0.4\textwidth]{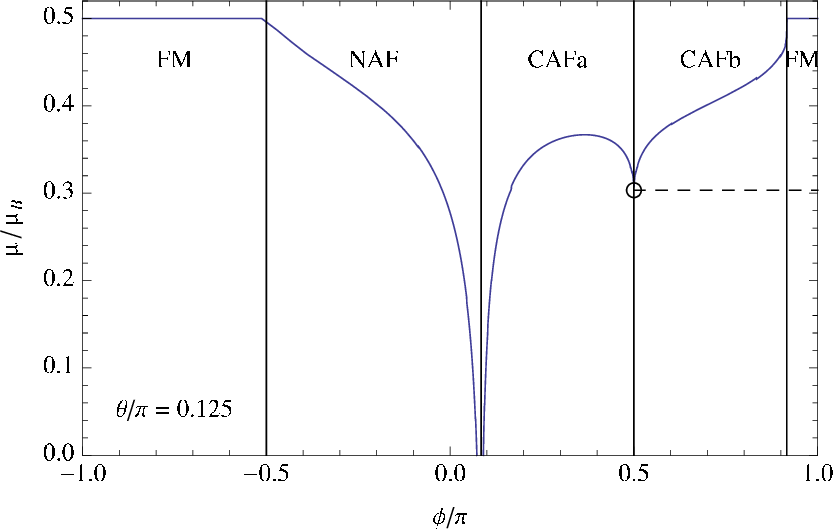}
    \includegraphics[width=0.4\textwidth]{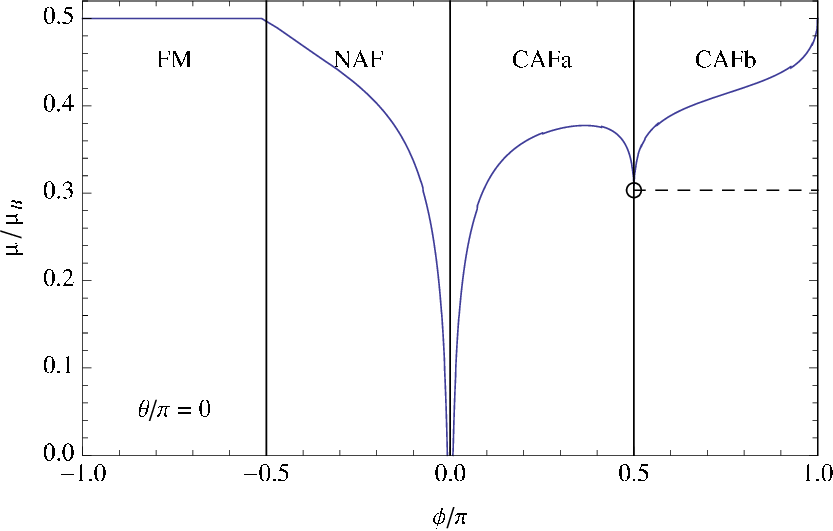}
    \caption{Ordered moment $m_{\text s}=\mu/\mu_{\text B}$ for fixed
    anisotropy parameter $\theta$ as a function of the frustration
    angle $\phi$.  Top: $\theta=\pi/4$ (isotropic case,
    $J_{1a}=J_{1b}$).  The ordered moment vanishes for
    $0.1150\lesssim\phi/\pi\lesssim0.1508$ (NAF--CAFa/b) and
    $0.8491\lesssim\phi/\pi\lesssim0.8524$ (CAFa/b--FM crossover).
    Middle: $\theta=\pi/8$.  The gap between NAF and CAFa is smaller
    but still finite, the disordered region between CAFb and FM has
    disappeared.  Bottom: $\theta=0$, corresponding to $J_{1b}=0$.
    The gap $m_{\text s}=0$ appears for
    $-0.00727\lesssim\phi/\pi\lesssim0.00775$.  The dashed horizontal
    line in the three plots denotes the value $m_{\text
    s}\approx0.3034$ obtained for the standard Heisenberg model with
    $J_{1a}=J_{1b}$ and $J_{2}=0$.}
    \label{fig:ms}
\end{figure}
Fig.~\ref{fig:ms} displays the behavior of $m_{\text s}=\mu/\mu_{\text
B}$ as a function of the frustration angle $\phi$ for three different
anisotropy parameters $\theta$.  The upper plot shows the isotropic
case (see also Ref.~\onlinecite{uhrig:09}), $\theta=\pi/4$, corresponding 
to $J_{1a}=J_{1b}=J_{1}$.  Coming
from the FM phase for $\phi<-\pi/2$, $m_{\text s}$ is gradually
suppressed to the well-known value $m_{\text s}^{0}\approx0.3034$ at
$\phi=0$ ($J_{2}=0$), corresponding to the NAF state of the 
nearest-neighbor Heisenberg model.

Increasing $\phi$ towards the NAF--CAFa/b boundary further reduces
$m_{\text s}$, until at $\phi/\pi\approx0.1150$ (where
$J_{2}/J_{1}\approx0.3779$), the ordered moment vanishes.  The
classical CAFa/b--NAF border is given by $J_{2}/J_{1}=1/2$, or
$\phi/\pi\approx0.1476$.  Soon after entering the CAFa/b regime, at
$\phi/\pi\approx0.1508$ or $J_{2}/J_{1}\approx0.5129$, $m_{\text s}$
becomes finite again and grows rapidly towards an almost constant
value $m_{\text s}\approx m_{\text s}(J_{1}=0)$.

At $J_{1}=0$ or $\phi=\pi/2$, the lattice can be subdivided into two
noninteracting sublattices with a nearest-neighbor interaction
$J=J_{2}$, therefore we must have $m_{\text s}(\phi=\pi/2)\equiv
m_{\text s}(\phi=0)$.  This is indicated by the dashed horizontal
line which illustrates that throughout most of the CAF region the
moment reduction is almost the same as that of the unfrustrated 
NAF. In fact for $\phi=\pi/2$ the CAF moment is {\it stabilized} by
quantum fluctuations which orient the moments of the two sublattices
parallel. This is the so-called `order by disorder' mechanism.

The behavior of the ordered moment in the CAFa/b phase is independent
of the sign of $J_{1}$ and therefore symmetric around $\phi=\pi/2$.
Reaching the border to the FM phase, $m_{\text s}$ sharply drops back
to zero at $\phi/\pi\approx0.8492$ or $J_{2}/J_{1}\approx-0.5129$.
The CAFa/b--FM border is given by $J_{2}/J_{1}=-1/2$, or
$\phi/\pi\approx0.8524$.  At this border, $m_{\text s}$ immediately
jumps to the saturation value $m_{\text s}=1/2$ in the FM phase.

Now we turn to the anisotropic case, $\theta\ne\pi/4$.  The lower two
plots of Fig.~\ref{fig:ms} show $m_{\text s}$ for $\theta=\pi/8$,
corresponding to $J_{1b}/J_{1a}=\sqrt{2}-1\approx0.41$, and the
fully anisotropic case $\theta=0$, meaning $J_{1b}=0$.  For
$\phi/\pi<1/2$, the overall behavior of $m_{\text s}$ is similar to
the isotropic case: After leaving the FM regime, $m_{\text s}$ is
suppressed down to zero, and becomes finite again after the crossover
to the CAFa phase.  However, there are two quantitative differences:
Firstly, the region where $m_{\text s}=0$ is smaller than for
$\theta=\pi/4$.  Secondly and most importantly inside the columnar AF phases, the ordered
moment is restored to even larger values than for isotropic exchange.

Exactly at $\phi=\pi/2$, we have $J_{1a}=J_{1b}=0$, and $m_{\text s}$
shows a dip with the universal value $m_{\text s}\approx0.3034$ of the
nearest-neighbor Heisenberg model, indicated by the dashed lines in
the plots.  For any value of $\theta$, at this point only $J_{2}$ is
finite, and the same argument as for the isotropic case applies; we
must have $m_{\text s}(\phi=\pi/2)\equiv m_{\text s}(\phi=0)$ for
arbitrary ratios $J_{1b}/J_{1a}$.

In contrast to the isotropic case, the symmetry around the point
$\phi=\pi/2$ is lost, and $m_{\text s}$ is restored in the CAFb phase
towards the saturation value upon entering the FM phase.  There is no
region around the CAFb/FM border where $m_{\text s}$ is suppressed as
in the isotropic case.

\begin{figure}
    \centering
    \includegraphics[width=0.4\textwidth]{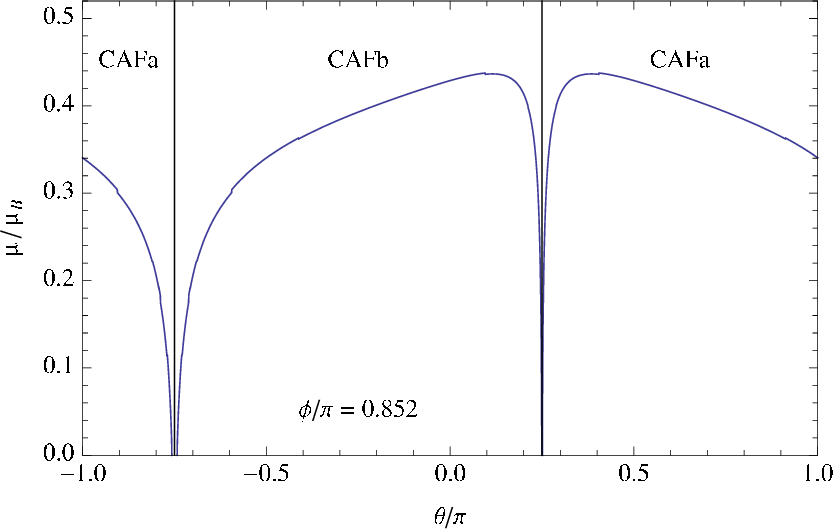}
    \caption{Ordered moment $m_{\text s}$ for $\phi/\pi=0.852$ as a
    function of the anisotropy parameter $\theta$. The frustration 
    angle is chosen such that in the isotropic case ($\theta=\pi/4$), 
    the system is in the disordered regime at the CAFa/b--FM corner.}
    \label{fig:mstheta}
\end{figure}
A vanishing ordered moment implies that, at least within our
approximation, the order parameter for the corresponding classical
phase is destroyed by quantum fluctuations.  Historically this was one
of the first indications of the appearance of an intermediate phase
without magnetic order, and our findings suggest that the well-known
disordered phase for the isotropic $J_{1}$-$J_{2}$ model for AF
exchange couplings extends to the whole range of anisotropic
interactions with arbitrary ratios $J_{1b}/J_{1a}$.

This is {\em not\/} the case for the CAFa/b--FM crossover, where
numerically already at a deviation $\Delta\theta/\theta<0.01$
from the isotropic value $\theta=\pi/4$ the ordered moment remains
well-defined around the classical CAFa/b--FM transition point.
Fig.~\ref{fig:mstheta} illustrates this behavior: The plot shows the
ordered moment as a function of the anisotropy parameter for fixed
frustration angle $\phi/\pi=0.8520$ in the whole phase diagram.  The
two sharp dips at $\theta=-3\pi/4$ and $\pi/4$ correspond to the
disordered regimes at the CAFa/b--NAF corner and the CAFa/b--FM corner
in the phase diagram, respectively.  At the CAFa/b--FM corner around
$\theta=\pi/4$, we have $m_{\text s}=0$ only for a tiny range
$0.2493\lesssim\theta/\pi\lesssim0.2507$.  However, the precise
relation to the extension of the disordered phase around this corner
for finite orthorhombic anisotropy remains unclear.\\

\begin{figure}
    \centering
    \includegraphics[width=0.4\textwidth]{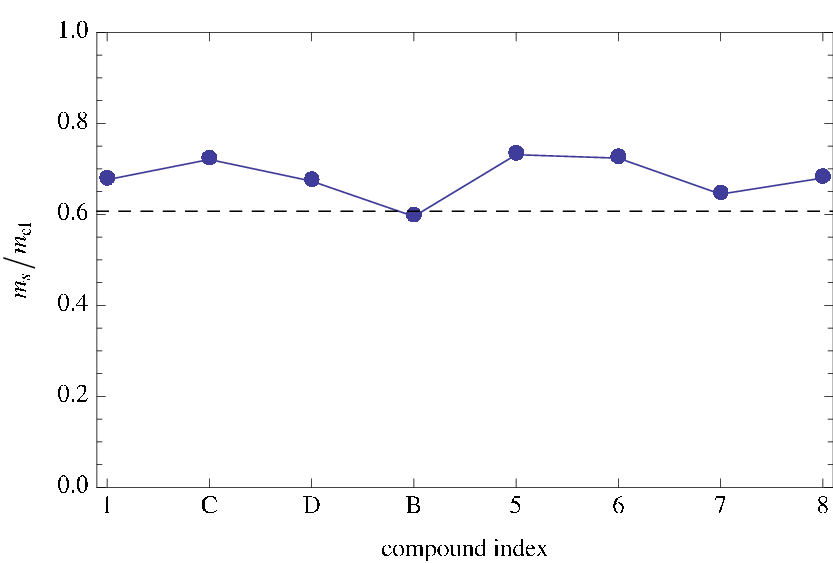}
    \caption{Ordered moments $m_{\text s}(\phi,\theta)$ normalized to
    the classical constant $m_{\text{cl}}=S$ for the compounds listed
    in Table~\ref{tbl:xchng} calculated with Eq.~(\ref{eqn:ms}).  The
    dashed horizontal line indicates $m_{\text
    s}(\phi=0,\theta=\pi/4)$ for the simple $J_{2}=0$, $J_{1a}=J_{1b}$
    nearest-neighbor Heisenberg model (Point A in 
    Fig.~\ref{fig:egslsw}).}
    \label{fig:msexp}
\end{figure}

In summary, if one regards the lower two panels in Fig.\ref{fig:ms}
representing the anisotropic case one observes a remarkable fact: The
moment reduction by quantum fluctuation in the CAFa/b phases is {\em
less\/} than in the {\em unfrustrated\/} simple nearest-neighbor NAF
phase (open circle), except for a very small region close to the
strongly frustrated CAFa/NAF boundary line.  If one compiles the
reduced moments $m_{\text s}(\phi,\theta)$ for the proposed parameter
sets of Fe-pnictide compounds (Table~\ref{tbl:xchng}) as shown in
Fig.~\ref{fig:msexp} it is obvious that in most cases the moment
reduction by quantum fluctuations for the proposed CAFa models is {\em
less\/} than in the simple nearest-neighbor NAF

This result is in part due to the stabilization of the moment due to
the effect of the anisotropy as visible from Fig.~\ref{fig:ms} which
extends the stable range of $\phi$ for the CAFa phase in
Fig.~\ref{fig:egslsw}.  In fact the frustration angle for
BaFe$_{2}$As$_{2}$ (D) is rather close to the strongly frustrated
value $\phi/\pi$=0.15 of the {\em isotropic} ($\theta/\pi=0.25$)
model; nevertheless it is at a considerable distance from the
anisotropic CAFa/NAF instability line and hence shows only moderate
moment reduction.  Therefore from Fig.~\ref{fig:msexp} we conclude
that frustration/quantum fluctuation effects within a local moment
picture may not be used to explain the surprisingly small ordered
moment of the pnictides.  We note, however that this conclusion does
not invalidate the usefulness of the $J_{1a,b}$-$J_2$ local moment
model for the interpretation of INS spin wave results.  In classical
(linear) spin wave theory only the products $SJ_i$ enter the spin wave
velocity and dispersion $SE(\vec k)$ and therefore the shape of the
dispersion does not depend on the size of the staggered moment as long
as this approximation is reasonable.

\section{Field dependent properties of the $J_{1a,b}$-$J_2$ model}

\label{sect:Magnetic}

It has been shown that high field magnetization (up to the saturation
field $h_{\text s}$) is an excellent way to investigate the isotropic
$J_1$-$J_2$ model~\cite{schmidt:07b,thalmeier:08} because the degree
of exchange frustration determines the characteristic nonlinearity of
the magnetization curve.  This is especially true for the parameter
region where the CAF phase is realized.  One should remark however
that in the undoped Fe-pnictides the energy scale of $J_{\text
c}\simeq 5\cdot 10^2\,\text K$ is too large to reach the high
field region.  However it might be feasible for some of the doped
Fe-pnictides where the ordering temperature $T_{\text m}$ and hence
$J_{\text c}$ are strongly reduced, provided that doping in the
spacing layers does not impede the usefulness of the $J_{1a,b}$-$J_2$
model for the spin excitations in the FeAs layers.

\subsection{Magnetization at high fields}

The total magnetization of the system is the ground-state expectation
value of the $z$ component of the spin $\vec S$ in the global
coordinate system,
\begin{equation}
    m=\frac{1}{N}\sum_{i}\left\langle S_{i}^{z}\right\rangle.
\end{equation}
Since this is just the projection of the ordered magnetic moments 
onto the field direction, we can also write
\begin{equation}
    m=m_{\text s}\cos\Theta.
\end{equation}
Here, $\Theta$ is {\em not\/} the classical canting angle
$\Theta_{\text c}$ [this would describe the field dependence of the
classical spin system, i.\,e., a straight line $m(h)=S(h/h_{\text
s})$], but rather must include the first-order corrections from linear
spinwave theory.  We thus have to regard $\Theta$ as independent
variable again, return to the Hamiltonian given by
Eq.~(\ref{eqn:hbog}) before the replacement $\Theta\to\Theta_{\text
c}$ and minimize its corresponding ground-state energy with respect to
$\Theta$.

Since $1/S$ corrections are already included in the ground-state
energy $E_{\text{gs}}(\Theta_{\text c})={\cal H}_{\text{cl}}+{\cal
H}_{\text{zp}}$ of the linear spin-wave Hamiltonian given by
Eqs.~(\ref{eqn:hcl0}), (\ref{eqn:hbog}), (\ref{eqn:hzp}),
and~(\ref{eqn:ek}), we can equivalently use the definition of the
total magnetization per site as the negative field derivative of
$E_{\text{gs}}(\Theta_{\text c})$,
\begin{equation}
    m=-\frac{1}{N}\frac{\partial}{\partial h}E_{\text{gs}}(\Theta_{\text c})
    \label{eqn:mdef}
\end{equation}
with $\Theta_{\text c}$ given by Eq.~(\ref{eqn:thetaca0}). The result 
is
\begin{equation}
    m=
    S
    \frac{h}{h_{\text s}}
    \left[
    1+\frac{1}{2S}\frac{1}{N}\sum_{\vec k}
    \frac{B(\vec k)\left(A(\vec k)-B(\vec k)\right)}{A(0)E(h,\vec k)}
    \right]
    \label{eqn:m}
\end{equation}
(see Appendix~\ref{app:m} for details).

\begin{figure}
    \centering
    \includegraphics[width=.4\textwidth]{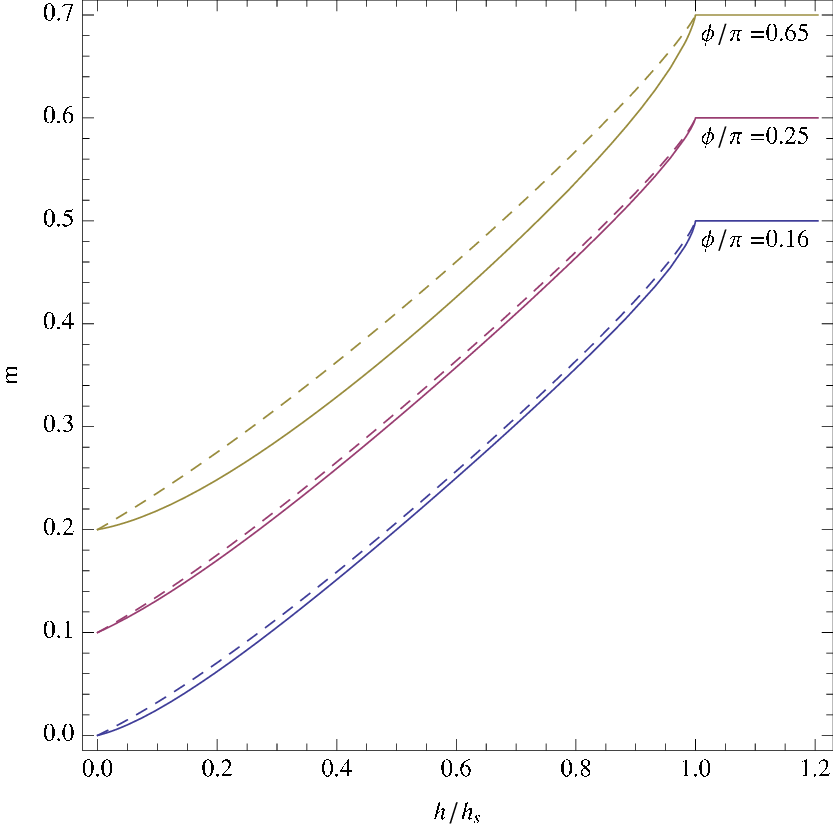}
    \caption{Uniform magnetic moment $m$ per site as a function of the applied
    magnetic field $h/h_{\text s}$ at three different frustration 
    angles in the CAF phases, $\phi/\pi=0.16$ (near NAF), $0.25$ 
    (CAFa), and $0.65$ (CAFb near FM).  Between each pair of
    adjacent curves an offset $\Delta m=0.1$ is inserted. The solid 
    lines denote the field dependence in the isotropic case, 
    $\theta=\pi/4$, the dashed lines denote the maximally 
    anisotropic case, $\theta=0$.}
    \label{fig:mone}
\end{figure}

Fig.~\ref{fig:mone} displays three sets of curves of $m(h)$ for
different frustration angles.  The magnetic field is normalized to the
respective saturation field.  The solid curves show the
field dependence of the induced moment $m$ in the isotropic case
$\theta=\pi/4$, $J_{1a}=J_{1b}$, and the dashes curves show the same
quantity in the maximally anisotropic case $\theta=0$ or $J_{1b}=0$.

Deep inside the ordered phases (well separated from phase boundaries), the
first-order corrections to the total moment are small, as indicated by
the small bending of $m(h)$ for $\phi/\pi=0.25$ (middle curve) in
Fig.~\ref{fig:mone}.  It is only near the crossover between adjacent
different phases, where corrections become strong, and the field
dependence of the magnetization differs significantly from the
classical behavior, as shown in $m(h)$ for $\phi/\pi=0.65$ with
isotropic exchange constants.

Introducing an anisotropy generally reduces the quantum corrections
leading to the nonlinear magnetization (dashed curves).  According to
Table~\ref{tbl:xchng} the Fe pnictides are not close to the phase
boundaries, therefore the effect of the anisotropy on the nonlinear
magnetization curves is not very prominent as indeed suggested by
Fig.~\ref{fig:mone}.

\subsection{Magnetic susceptibility at low fields}

Another sensitive probe to the degree of frustration is the low field
uniform susceptibility $\chi=\partial m/\partial h$  which may
be used to obtain further insight as an alternative to the ordered moment.
We  obtain

\begin{widetext}
\begin{equation}
    \chi=
    \frac{1}{2A(0)}\left[
    1+\frac{1}{2S}\frac{1}{N}\sum_{\vec k}
    \frac{B(\vec k)\left(A(\vec k)-B(\vec k)\right)}%
    {A(0)E(h,\vec k)}   
    +\frac{1}{S}\cos^{2}\Theta_{\text c}\frac{1}{N}\sum_{\vec k}
    \frac{B^{2}(\vec k)\left(A(\vec k)-B(\vec k)\right)^{2}}%
    {A(0)E^{3}(h,\vec k)}
    \right]
    \label{eqn:chi}
\end{equation}
\end{widetext}
where the terms in brackets include the first-order corrections to the 
classical value,
\begin{equation}
    \chi_{\text{cl}}=\frac{1}{2A(0)}
    =\frac{S}{h_{\text s}}
    =\mbox{const.}
\end{equation}
In the FM regions, $\chi$ is undefined, since the system at $T=0$ is
already fully polarized.  In the three AF phases, $\chi$ diverges near
the FM phase, apart from those areas, where already the vanishing
ordered moment discussed above indicates that the ordered phase is no
longer stable.

\begin{figure}
    \centering
    \includegraphics[width=.4\textwidth]{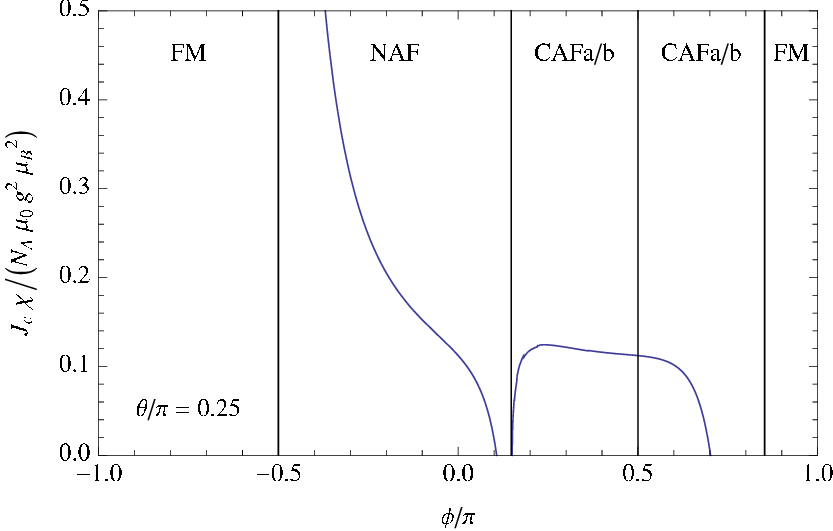}
    \includegraphics[width=.4\textwidth]{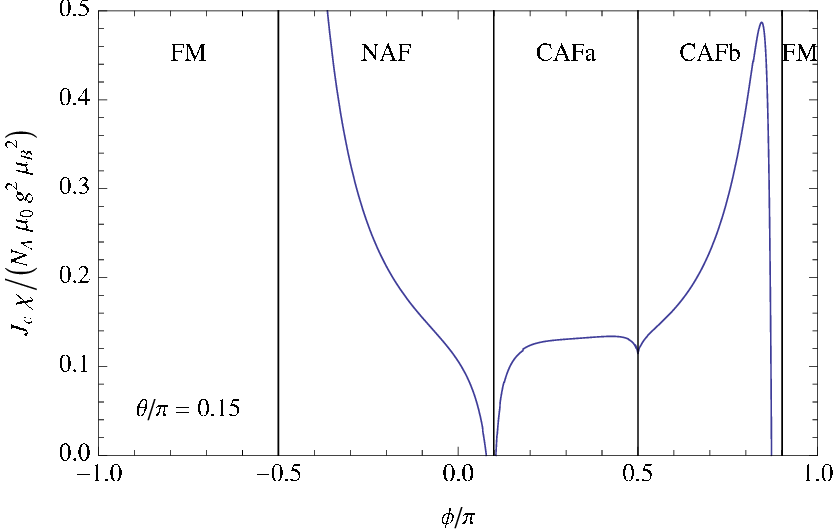}
    \includegraphics[width=.4\textwidth]{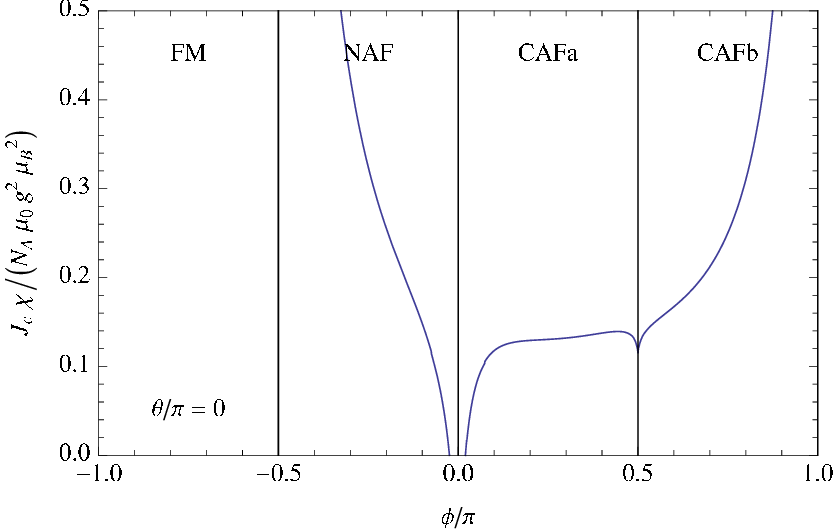}
    \caption{Uniform magnetic susceptibility $\chi=\left(\partial
    m/\partial h\right)_{h\to0}$ as a function of the frustration
    angle $\phi$ for (from top to bottom) $\theta/\pi=1/4$,
    $\theta/\pi=0.15$, and $\theta=0$.}
    \label{fig:chione}
\end{figure}
Fig.~\ref{fig:chione} shows the dependence of $\chi$ on the
frustration angle $\phi$ for three different anisotropy parameters
$\theta$.  In the top of Fig.~\ref{fig:chione}, the isotropic case is
shown.  The susceptibility diverges at the crossover from the NAF to
the FM phase ($J_{1a}=J_{1b}=0$, $\phi/\pi=-1/2$).  Around the border
to the CAF phases ($\phi=\tan^{-1}(1/2)$), it vanishes and becomes
undefined.  Qualitatively the same happens near the border to the FM
phase at $\phi=\pi-\tan^{-1}(1/2)$.  Apart from shifting the classical
phase boundaries, introducing an orthorhombic anisotropy generally
stabilizes the magnetic CAF ground state.  Therefore the `gaps' where
strong frustration destroys the magnetic order are gradually closed,
see the center plot of Fig.~\ref{fig:chione}.

At the CAFb/FM crossover, the behavior of $\chi$ even changes to a
divergence upon increasing the orthorhombic asymmetry by lowering
$\theta$.  The bottom part of Fig.~\ref{fig:chione} shows $\chi$ for
$J_{1b}=0$, where the ``gap'' to the FM phase is closed.  This is
fully compatible with the rapid stabilization of the CAFb ordered
moment as function of increasing anisotropy (see.  Fig~\ref{fig:ms})
at the boundary to FM.

At $\phi=\pi/2$, the $\phi$ dependence of the magnetic susceptibility
has the same feature as the ordered moment discussed in
Sec.~\ref{sect:ms}: At $\phi=\pi/2$, $J_{1}=0$, and we must have
$\chi(\phi=0,\theta=\pi/4)\equiv\chi(\phi=\pi/2,\theta=\text{arbitrary})$,
therefore a small dip appears in the orthorhombic case.

In the CAFa sector relevant for the pnictides the susceptibility has
a plateau value except very close to the gap of instability. The 
value is almost equal to that for the unfrustrated simple
n.n. NAF. This underlines again that quantum fluctuations
due to frustration cannot explain the anomalous magnetism of Fe pnictides.
 
\begin{figure}
    \centering
    \includegraphics[width=.4\textwidth]{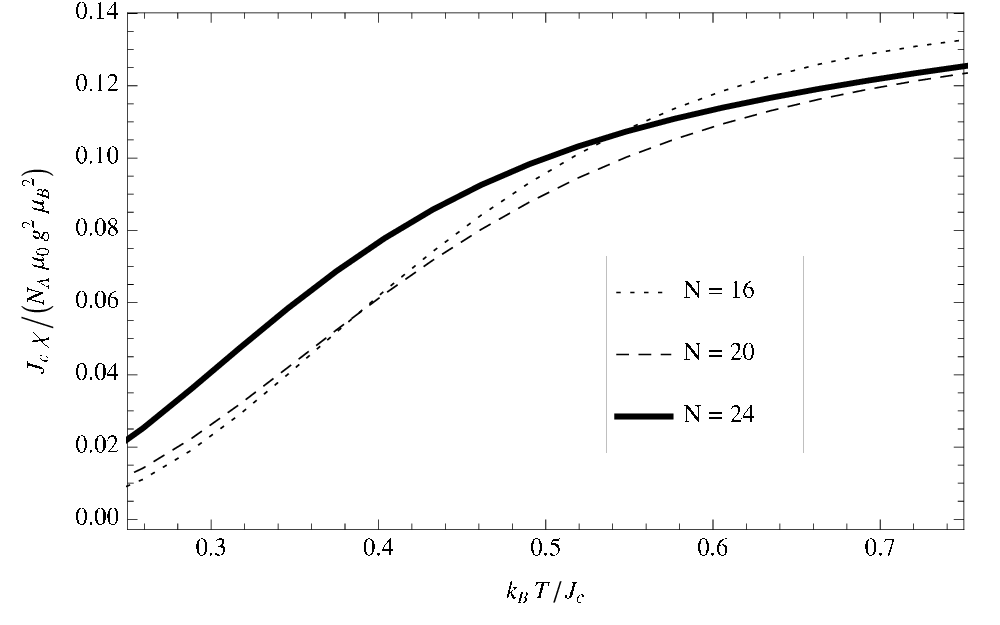}
      \caption{Temperature dependence of the uniform magnetic
      susceptibility for $\phi/\pi=0.35$, and $\theta=0.25$ as for (B)
      in Table~\ref{tbl:xchng} with $J_{\text c}=58.5\,\text{meV}$,
      corresponding to a temperature variation between $170\,\text K$
      and $510\,\text K$.}
    \label{fig:chiT}
\end{figure}
Furthermore the uniform susceptibility was found to have an unexpected
temperature dependence.  Within the (semi-)metallic itinerant model
characterized by electron and hole pockets one would naively expect a
constant Pauli susceptibility above the ordering temperature $T_{\text
m}$ and a reduction below due to the gap opening.  While the latter
was found for numerous pnictide compounds, the susceptibility above
$T_{\text m}$ is not constant but still increases roughly linearly
with temperature \cite{klingeler:09}.  Explanations for this observation were given within
the noninteracting two band model \cite{sales:09}, an interacting
Fermi liquid picture including nonanalytic correction terms
\cite{korshunov:09}, within a model of coexisting itinerant and
localized moments \cite{kou:09}.

On the other hand INS results have shown that low energy spin
excitations can be well described by a suitably parametrized
$J_{1a,b}$-$J_2$ local moment model according to
Table~\ref{tbl:xchng}. Then one should expect something similar for
the low (zero) frequency susceptibility, at least qualitatively.  To
check this conjecture we performed finite temperature Lanczos
calculations as, e.g., described in Ref.~\onlinecite{shannon:04} for
various finite $J_{1a,b}$-$J_2$ clusters.  The result is shown in
Fig.~\ref{fig:chiT} for a parameter set (B) corresponding to
BaFe$_2$As$_2$.  For temperatures $T\lesssim0.25J_{\text c}$ the
calculation becomes unreliable due to finite size effects.  The
temperature range corresponds to $\sim170\,\text K$ at the lower and
$\sim 510\,\text K$ at the upper boundary.  The increase linearity in
$T$ observed for BaFe$_2$As$_2$ from $150\,\text K$ to $300\,\text K$
is qualitatively reproduced although the absolute increase is too
large.  We mention that in the combined itinerant-localized model of
Ref.~\onlinecite{kou:09} the increased linear $T$-dependence was also
attributed to the local moment contribution.  At the very least our
calculation shows clearly that in the present range of measurement one
should {\em not\/} yet expect the high temperature Curie law
$\chi(T)\sim1/T$ for local moment systems.  This should be expected
only quite above the maximum temperature for $\chi(T)$, which is at
about $1.1J_{\text c}$ or $750\,\text K$ in the case of
Fig.~\ref{fig:chiT}.

\section{Summary and Conclusion}
\label{sect:Conclusion}

The local moment model for Fe pnictides has been surprisingly useful
to explain the low energy spin excitations obtained in INS
phenomenologically, albeit with the assumption of possibly very
anisotropic exchange.  The latter may have its microscopic origin in
underlying orbital order as proposed in \cite{kubo:09,zhou:09} but
this is still unexplored.  The very usefulness of the local moment
picture may be a consequence of Hund's rule correlations in the
multiorbital state of Fe pnictides \cite{zhou:09}.

In this work we have investigated in detail the empirical localized
moment $J_{1a,b}$-$J_2$ model, in particular the effect of the
in-plane anisotropy and the frustration effect.  It has been a
recurrent topic to explain the comparatively small ordered moments in
Fe-pnictides as the effect of enhanced quantum fluctuations in the
ground state due to large degeneracy caused by frustrated $J_{1a,b}$
and $J_2$ exchange bonds.

We have investigated this question in detail using spinwave
approximation and in part the exact-diagonalization Lanczos method to
calculate ground state energy, phase diagram and moment reduction by
quantum fluctuation as function of anisotropy and frustration
parameters.  In addition we have studied high field magnetization and
low field uniform susceptibility.

We found that generally the anisotropy lifts the degeneracy between
CAFa/b phases and extends their stability range as a function of
frustration.  Furthermore the anisotropy reduces or closes the
instability gap on the phase boundary to the NAF or FM phase
respectively.  Most importantly we have shown that in the CAFa sector
relevant for the pnictides according to Table~\ref{tbl:xchng} the
moment reduction by quantum fluctuations is generally {\em less\/}
than for the simple unfrustrated n.n. N\'eel antiferromagnet.  The
same result can be obtained from the uniform low field susceptibility.

Therefore we conclude that the anomalously low moment in the pnictides
is not explained by quantum fluctuations in effective localized moment
models but needs a more microscopic viewpoint including the itinerant
multiorbital nature of the magnetic state.  Such proposals have been
made within recent ab-initio calculations using the full orbital basis
\cite{cricchio:09,lee:09,zhang:10}.  This does not invalidate,
however, the exceptional usefulness of the simple $J_{1a,b}$-$J_2$
model to describe the low energy spin excitations.

We have also derived the spin wave excitations in an external field
for the anisotropic model.  It remains to be seen whether new
information on the exchange models can be gained from INS experiments
in finite fields.

Finally, in a corollary we address a result of our analysis not
immediately relevant for pnictides because it is related to the
magnetic instability at the CAFa/b--FM boundary ($\phi/\pi=0.852$).
There are 2D local moment compounds \cite{thalmeier:08} where the
frustration angle is quite close to that boundary, contrary to the Fe
pnictides.  It has been shown for the isotropic model that the true
ground state in this region is of the spin-nematic hidden order state
\cite{shannon:06}.  Although spin wave theory is not adequate to fully
address this question we have shown (Fig.~\ref{fig:mstheta}) that the
columnar order at the boundary recovers immediately when turning on even
a tiny anisotropy of n.n. exchange constants $J_{1a,b}$.  Since small
anisotropies usually exist in such compounds we predict that the spin
nematic state of the isotropic $J_1$-$J_2$ model will be very hard to
find in a real compound.


\appendix
    
\section{Linear spin-wave analysis}
\label{app:hp}

The formal procedures described here closely follow and generalize
those discussed in Refs.~\onlinecite{ohyama:94}
and~\onlinecite{veillette:05}, where linear spin-wave theory has been
applied to the triangular lattice. In the sections of this appendix,
we don't make any assumptions about lattice geometry, dimensionality,
and exchange constants except from the requirement that U(1) symmetry
is conserved and still exists upon switching on a magnetic field.  The
Hamiltonian is assumed to have the general form given by
Eqs.~(\ref{eqn:ham}) and~(\ref{eqn:jij}).

Dropping the primes (working in the local coordinate system), we use
boson operators $a_{i}$ and $a_{i}^{\dagger}$ and write
\begin{eqnarray*}
    S_{i}^{z}&=&S-a_{i}^{\dagger}a_{i},
    \\
    S_{i}^{+}&=&\sqrt{2S}
    \left(1-\frac{a_{i}^{\dagger}a_{i}}{2S}\right)^{1/2}a_{i}
    \to\sqrt{2S}a_{i},
    \\
    S_{i}^{-}&=&\sqrt{2S}a_{i}^{\dagger}
    \left(1-\frac{a_{i}^{\dagger}a_{i}}{2S}\right)^{1/2}
    \to\sqrt{2S}a_{i}^{\dagger},
    \\
    S_{i}^{x}&=&\frac{1}{2}\left(S_{i}^{+}+S_{i}^{-}\right)
    \to\sqrt{\frac{S}{2}}
    \left(a_{i}+a_{i}^{\dagger}\right),
    \\
    S_{i}^{y}&=&\frac{1}{2{\rm i}}\left(S_{i}^{+}-S_{i}^{-}\right)
    \to-{\rm i}\sqrt{\frac{S}{2}}
    \left(a_{i}-a_{i}^{\dagger}\right).
\end{eqnarray*}
Keeping only terms up to bilinear order in the boson operators, we
expand the scalar products in Eq.~(\ref{eqn:ham}).  The Hamiltonian up
to bilinear order then reads
\begin{widetext}
\begin{eqnarray*}
    \lefteqn{{\cal H}\to{\cal H}_{\text{cl}}}
    \nonumber\\&&{}
    +\frac{S}{2}\sum_{\langle ij\rangle}
    \left[
    \left(a_{i}^{\dagger}a_{j}+a_{i}a_{j}^{\dagger}\right)
    \left(
    J_{ij}^{\perp}\cos(\vec Q\vec R_{ij})
    \left(\cos^{2}\Theta+1\right)
    +J_{ij}^{z}\sin^{2}\Theta
    \right)
    \right.
    \nonumber\\&&
    \phantom{\frac{S}{2}\sum_{\langle ij\rangle}}
    +\left(a_{i}a_{j}+a_{i}^{\dagger}a_{j}^{\dagger}\right)
    \left(
    J_{ij}^{\perp}\cos(\vec Q\vec R_{ij})
    \left(\cos^{2}\Theta-1\right)
    +J_{ij}^{z}\sin^{2}\Theta
    \right)
    \nonumber\\&&
    \phantom{\frac{S}{2}\sum_{\langle ij\rangle}}
    -2\left(a_{i}^{\dagger}a_{i}+a_{j}^{\dagger}a_{j}\right)
    \left(
    J_{ij}^{\perp}\cos(\vec Q\vec R_{ij})\sin^{2}\Theta
    +J_{ij}^{z}\cos^{2}\Theta
    \right)
    \nonumber\\&&
    \phantom{\frac{S}{2}\sum_{\langle ij\rangle}}
    -2{\rm i}\left(a_{i}^{\dagger}a_{j}-a_{i}a_{j}^{\dagger}\right)
    J_{ij}^{\perp}\sin(\vec Q\vec R_{ij})\cos\Theta
    \nonumber\\&&
    \phantom{\frac{S}{2}\sum_{\langle ij\rangle}}
    -{\rm i}\sqrt{2S}\left(
    a_{i}-a_{i}^{\dagger}-a_{j}+a_{j}^{\dagger}
    \right)
    J_{ij}^{\perp}\sin(\vec Q\vec R_{ij})\sin\Theta
    \nonumber\\&&{}
    \left.
    \phantom{\frac{S}{2}\sum_{\langle ij\rangle}}
    +\sqrt{2S}\left(
    a_{i}+a_{i}^{\dagger}+a_{j}+a_{j}^{\dagger}
    \right)
    \left(J_{ij}^{z}-J_{ij}^{\perp}\cos(\vec Q\vec R_{ij})\right)
    \cos\Theta\sin\Theta
    \right]
    \nonumber\\&&{}
    +h\sum_{i}
    \left[
    a_{i}^{\dagger}a_{i}\cos\Theta-
    \sqrt{\frac{S}{2}}\left(a_{i}+a_{i}^{\dagger}\right)\sin\Theta
    \right].
\end{eqnarray*}
\end{widetext}
In the above sum, the contribution
\[
    \frac{S}{2}\sum_{\langle ij\rangle}\left[
    -{\rm i}\sqrt{2S}\left(
    a_{i}-a_{i}^{\dagger}-a_{j}+a_{j}^{\dagger}
    \right)
    J_{ij}^{\perp}\sin(\vec Q\vec R_{ij})\sin\Theta
    \right]
\]
is antisymmetric in the variables $i$ and $j$ (site indices) and 
therefore vanishes upon summation.

Inserting a Fourier representation of the spin wave operators
$a_{i}^{\dagger}$ into the above equation, replacing the sum over
bonds $\langle ij\rangle$ with a sum over sites $i$ plus their
neighbors $n$ gives after some rearrangement
\begin{widetext}
\begin{eqnarray*}
    \lefteqn{{\cal H}-{\cal H}_{\text{cl}}=}
    \nonumber\\&&
    \sqrt{2NS}\sin\Theta\left[
    S\cos\Theta
    \frac{1}{2}\sum_{n}
    \left(
    J_{n}^{z}-J_{n}^{\perp}\cos(\vec Q\vec R_{n})
    \right)
    -\frac{h}{2}
    \right]
    \left(a_{0}+a_{0}^{\dagger}\right)
    \nonumber\\&&
    +\frac{S}{2}\sum_{\vec k}\frac{1}{2}\sum_{n}
    e^{-{\rm i}\vec k\vec R_{n}}
    \left\{
    \left[
    J_{n}^{z}+J_{n}^{\perp}\cos(\vec Q\vec R_{n})
    -\cos^{2}\Theta
    \left(
    J_{n}^{z}-J_{n}^{\perp}\cos(\vec Q\vec R_{n})
    \right)
    \right]
    \left(
    a_{\vec k}^{\dagger}a_{\vec k}+a_{-\vec k}a_{-\vec k}^{\dagger}
    \right)
    \right.
    \nonumber\\&&
    \phantom{\frac{S}{2}\sum_{\vec k}\frac{1}{2}\sum_{n}
    e^{-{\rm i}\vec k\vec R_{n}}}
    +
    \left[
    J_{n}^{z}-J_{n}^{\perp}\cos(\vec Q\vec R_{n})
    -\cos^{2}\Theta
    \left(
    J_{n}^{z}-J_{n}^{\perp}\cos(\vec Q\vec R_{n})
    \right)
    \right]
    \left(
    a_{\vec k}a_{-\vec k}+a_{\vec k}^{\dagger}a_{-\vec k}^{\dagger}
    \right)
    \nonumber\\&&
    \phantom{\frac{S}{2}\sum_{\vec k}\frac{1}{2}\sum_{n}
    e^{-{\rm i}\vec k\vec R_{n}}}
    \left.
    +\frac{2}{{\rm i}}J_{n}^{\perp}\sin(\vec Q\vec R_{n})\cos\Theta
    \left(
    a_{\vec k}^{\dagger}a_{\vec k}-a_{-\vec k}a_{-\vec k}^{\dagger}
    \right)
    \right\}
    \nonumber\\&&
    +\frac{S}{2}\sum_{\vec k}\frac{1}{2}\sum_{n}
    \left[
    -2J_{n}^{\perp}\cos(\vec Q\vec R_{n})
    -2\cos^{2}\Theta
    \left(
    J_{n}^{z}-J_{n}^{\perp}\cos(\vec Q\vec R_{n})
    \right)
    \right]
    \left(
    a_{\vec k}^{\dagger}a_{\vec k}+a_{\vec k}a_{\vec k}^{\dagger}-1
    \right)
    \nonumber\\&&
    +\frac{h}{2}\cos\Theta\sum_{\vec k}
    \left(
    a_{\vec k}^{\dagger}a_{\vec k}
    +a_{\vec k}a_{\vec k}^{\dagger}-1
    \right).
\end{eqnarray*}
\end{widetext}
Performing the sum over the neighbors $n$, together with 
Eqs.~(\ref{eqn:jkdef}), (\ref{eqn:ak}), (\ref{eqn:bk}), 
and~(\ref{eqn:ck}) eventually leads to Eq.~(\ref{eqn:hlswb}).

We define
\[
    \hat a_{\vec k}^{\dagger}=\left(a_{\vec k}^{\dagger},a_{-\vec k}\right)
\]
to write
\begin{eqnarray*}
    {\cal H}&=&
    NS(S+1)\left(J_{\perp}(\vec Q)+A(0)\cos^{2}\Theta\right)
    \nonumber\\&& 
    -\frac{1}{2}Nh\left(2S+1\right)\cos\Theta
    +\frac{S}{2}\sum_{\vec k}
    \hat a_{\vec k}^{\dagger}
    H_{\vec k}
    \hat a_{\vec k},
    \\
    H_{\vec k}&=&
    \left(
    \begin{array}{cc}
	H_{1}+H_{\text a}
	&
	H_{2}
	\\
	H_{2}
	&
	H_{1}-H_{\text a}
    \end{array}
    \right),
    \\
    H_{1}&=&
    A(\vec k)
    -\cos^{2}\Theta\left(B(\vec k)+2A(0)\right)
    +\frac{h}{S}\cos\Theta,
    \\
    H_{2}&=&
    B(\vec k)\left(1-\cos^{2}\Theta_{\text c}\right),
    \\
    H_{\text a}&=&C(\vec k)\cos\Theta,
\end{eqnarray*}
dropping the part linear in $\{a_{\vec k=0}^{\dagger},a_{\vec k=0}\}$.
Since the operators $a_{\vec k}$ are bosons, their commutation
relations can be written as $\hat a_{\vec k}^{\dagger}\sigma_{z}\hat
a_{\vec k}=1$ where $\sigma_{z}$ is the symplectic unit matrix (which
in our case is identical to the $z$ Pauli spin matrix).  Assume
$U_{\vec k}$ is the matrix which diagonalizes Hamiltonian.
From the requirement that this transformation respects the
canonical commutation relations, it follows that $U_{\vec k}$ must be
symplectic, too,
\[
    U_{\vec k}^{\dagger}\sigma_{z}U_{\vec k}=\sigma_{z},
\]
and we have
\[
    U_{\vec k}^{\dagger}\sigma_{z}H_{\vec k}U_{\vec k}=\sigma_{z}D,
\]
where $D$ is the diagonal form of $H_{\vec k}$.  From this
transformation, we get the eigenvalues of
\[
    \sigma_{z}H_{\vec k}=\left(
    \begin{array}{cc}
	H_{1}+H_{\text a}
	&
	H_{2}
	\\
	-H_{2}
	&
	-H_{1}+H_{\text a}
    \end{array}
    \right)
\]
by evaluating the characteristic polynomial 
$\chi(E)=\det\left(\sigma_{z}H_{\vec k}-E\right)$. This polynomial 
can always be written as
\[
    \chi(E)=\det\left[\left(E-H_{\text a}\right)^{2}-
    \left(H_{1}-H_{2}\right)
    \left(H_{1}+H_{2}\right)
    \right],
\]
and the desired spin-wave dispersion can be immediately read off,
\begin{widetext}
\begin{eqnarray*}
    E(h,\vec k)&=&
    \sqrt{
    \left(H_{1}-H_{2}\right)
    \left(H_{1}+H_{2}\right)
    }+H_{\text a}
    \nonumber\\&=&
    \left\{
    \left[
    A(\vec k)-B(\vec k)-2A(0)\cos^{2}\Theta+\frac{h}{S}\cos\Theta
    \right]
    \right.
    \nonumber\\&&
    \left.
    \times\left[
    A(\vec k)+B(\vec k)\left(1-2\cos^{2}\Theta\right)
    -2A(0)\cos^{2}\Theta+\frac{h}{S}\cos\Theta
    \right]
    \right\}^{1/2}
    +C(\vec k)\cos\Theta 
    \nonumber\\&=&
    \sqrt{
    \left[
    A(\vec k)
    -\cos^{2}\Theta\left(B(\vec k)+2A(0)\right)
    +\frac{h}{S}\cos\Theta
    \right]^{2}
    -\left[
    B(\vec k)\left(1-\cos^{2}\Theta\right)
    \right]^{2}}
    +C(\vec k)\cos\Theta,
    \label{eqn:egsgeneral}
\end{eqnarray*}
\end{widetext}
where we have taken the root with the positive sign only.  Setting
$\Theta=\Theta_{\text c}$ eventually leads to the
dispersion~(\ref{eqn:ek}).

\section{Ordered moment}
\label{app:ms}

Writing the Hamiltonian, Eq.~(\ref{eqn:hlswb}), with the spin wave
operators $\alpha_{\vec k}^{\dagger}$, Eqs.~(\ref{eqn:alpha})
and~(\ref{eqn:alphad}), and diagonalizing the LSW Hamiltonian in a more explicit
way we get
\begin{widetext}
\begin{eqnarray*}
    u_{\vec k}&=&\mathop{\rm sign}\nolimits B(\vec k)\sqrt{
    \frac{1}{2}\left(
    \frac{A(\vec k)
    -\cos^{2}\Theta\left(B(\vec k)+2A(0)\right)
    +\frac{h}{S}\cos\Theta}{E(h,\vec k)}
    +1\right)},
    \\
    v_{\vec k}&=&\sqrt{
    \frac{1}{2}\left(
    \frac{A(\vec k)
    -\cos^{2}\Theta\left(B(\vec k)+2A(0)\right)
    +\frac{h}{S}\cos\Theta}{E(h,\vec k)}
    -1\right)}.
\end{eqnarray*}
\end{widetext}
Setting $\Theta=\Theta_{\text c}$ and inserting the coefficients
$v_{\vec k}$ into Eq.~(\ref{eqn:msgen}) yields the
expression~(\ref{eqn:ms}) for the ordered moment.

\section{Uniform moment and susceptibility}
\label{app:m}

The uniform moment is given by Eq.~(\ref{eqn:mdef}), or equivalently
\begin{equation}
    m
    =-\frac{1}{N}\frac{\partial E_{\text{gs}}(\Theta_{\text c})}%
    {\partial\cos\Theta_{\text c}}
    \frac{\partial\cos\Theta_{\text c}}{\partial h}
    \label{eqn:mm}
\end{equation}
with the ground-state energy in linear spin-wave approximation
\begin{eqnarray*}
    E_{\text{gs}}(\Theta_{\text c})&=&
    NS^{2}\left(J_{\perp}(\vec Q)-A(0)\cos^{2}\Theta_{\text c}\right)
    \nonumber\\&&{}
    +NSJ_\perp(\vec Q)
    +\frac{S}{2}\sum_{\vec k}E(h,\vec k),
\end{eqnarray*}
and $E(h,\vec k)$ given by Eq.~(\ref{eqn:ek}).
We have
\begin{eqnarray*}
    \frac{\partial E_{\text{gs}}(\Theta_{\text c})}%
    {\partial\cos\Theta_{\text c}}
    &=&
    -2A(0)\cos\Theta_{\text c}
    +\frac{S}{2}\sum_{\vec k}
    \frac{\partial E(h,\vec k)}%
    {\partial\cos\Theta_{\text c}},
    \nonumber\\
    \frac{\partial\cos\Theta_{\text c}}{\partial h}&=&
    \frac{1}{2SA(0)},
    \\
    \frac{\partial E(h,\vec k)}%
    {\partial\cos\Theta_{\text c}}
    &=&
    -2\cos\Theta_{\text c}
    \frac{B(\vec k)\left(A(\vec k)-B(\vec k)\right)}%
    {E(h,\vec k)},
\end{eqnarray*}
Inserting these expressions into Eq.~(\ref{eqn:mm}) gives the 
desired result, Eq.~(\ref{eqn:m}).

The correction to the magnetization (the deviation from the classical
behavior $m_{\text{cl}}=S\cos\Theta_{\text c}=S(h/h_{\text s})$) at this
level of approximation is a consequence of the zero-point fluctuations
only.  Integrating Eq.~(\ref{eqn:mdef}) between $h=0$ and $h=h_{\text
s}$, we can write
\begin{equation}
    \Delta E_{\text{gs}}
    =E_{\text{gs}}(h=h_{\text s})-E_{\text{gs}}(h=0)
    =-N\int_{0}^{h_{\text s}}m(h){\rm d}h.
    \label{eqn:diff}
\end{equation}
As discussed elsewhere, the zero-point fluctuations reduce
$E_{\text{gs}}(h=0)$, but do not reduce $E_{\text{gs}}(h_{\text s})$.
For the latter, the zero-point fluctuations vanish, again because the
fully polarized state at $h_{\text s}$ is an eigenstate of $\cal H$.
Thus the classical energy difference is smaller than the energy
difference including first-order corrections in Eq.~(\ref{eqn:diff}),
\begin{eqnarray}
    \lefteqn{\Delta E_{\text{gs}}^{\text{cl}}
    -\Delta E_{\text{gs}}^{\text{LSW}}}
    \nonumber\\
    &=&
    E_{\text{gs}}^{\text{cl}}(h_{\text s})
    -E_{\text{gs}}^{\text{cl}}(0)
    -\left(
    E_{\text{gs}}^{\text{LSW}}(h_{\text s})
    -E_{\text{gs}}^{\text{LSW}}(0)
    \right)
    \nonumber\\&=&
    E_{\text{gs}}^{\text{LSW}}(0)
    -E_{\text{gs}}^{\text{cl}}(0)
    \nonumber\\&=&
    NSJ_{\perp}(\vec Q)
    +\frac{S}{2}\sum_{\vec k}E(\vec k)
    \nonumber\\&=&
    E_{\text{zp}}
    \le0,
    \nonumber
\end{eqnarray}
hence the corrections to the integrated magnetization must be
negative.  Assuming a monotonic behavior of $m(h)$, the same relation
holds for the kernel in Eq.~(\ref{eqn:diff}), and it follows that
\[
    m_{\text{LSW}}(h)\le m_{\text{cl}}(h),\quad h\le h_{\text s},
\]
consequently we can expect that the corrected magnetization curve lies below
the classical one.

For the susceptibility, we differentiate $m$ once more with respect to 
the applied field,
\[
    \chi
    =
    \frac{\partial m}{\partial\cos\Theta_{\text c}}
    \frac{\partial\cos\Theta_{\text c}}{\partial h},
\]
and using
\begin{widetext}
\begin{eqnarray*}
    \frac{\partial m}{\partial\cos\Theta_{\text c}}
    &=&
    \frac{m}{\cos\Theta_{\text c}}
    +S\cos\Theta_{\text c}\frac{1}{2S}\frac{1}{N}\sum_{\vec k}
    \left[
    -\frac{B(\vec k)\left(A(\vec k)-B(\vec k)\right)}%
    {A(0)E^{2}(h,\vec k)}
    \frac{\partial E(h,\vec k)}{\partial\cos\Theta_{\text c}}
    \right]
    \nonumber\\
    &=&
    S\left[
    1+\frac{1}{2S}\frac{1}{N}\sum_{\vec k}
    \frac{B(\vec k)\left(A(\vec k)-B(\vec k)\right)}%
    {A(0)E(h,\vec k)}   
    +\cos^{2}\Theta_{\text c}\frac{1}{S}\frac{1}{N}\sum_{\vec k}
    \frac{B^{2}(\vec k)\left(A(\vec k)-B(\vec k)\right)^{2}}%
    {A(0)E^{3}(h,\vec k)}
    \right].
    \nonumber
\end{eqnarray*}
\end{widetext}
We obtain Eq.~(\ref{eqn:chi}).

\bibliography{Fepnic}

\begin{thebibliography}{10}%
\makeatletter
\providecommand \@ifxundefined [1]{%
 \ifx #1\undefined \expandafter \@firstoftwo
 \else \expandafter \@secondoftwo
\fi
}%
\providecommand \@ifnum [1]{%
 \ifnum #1\expandafter \@firstoftwo
 \else \expandafter \@secondoftwo
\fi
}%
\providecommand \enquote [1]{``#1''}%
\providecommand \bibnamefont  [1]{#1}%
\providecommand \bibfnamefont [1]{#1}%
\providecommand \citenamefont [1]{#1}%
\providecommand\href[0]{\@sanitize\@href}%
\providecommand\@href[1]{\endgroup\@@startlink{#1}\endgroup\@@href}%
\providecommand\@@href[1]{#1\@@endlink}%
\providecommand \@sanitize [0]{\begingroup\catcode`\&12\catcode`\#12\relax}%
\@ifxundefined \pdfoutput {\@firstoftwo}{%
 \@ifnum{\z@=\pdfoutput}{\@firstoftwo}{\@secondoftwo}%
}{%
 \providecommand\@@startlink[1]{\leavevmode\special{html:<a href="#1">}}%
 \providecommand\@@endlink[0]{\special{html:</a>}}%
}{%
 \providecommand\@@startlink[1]{%
  \leavevmode
  \pdfstartlink
   attr{/Border[0 0 1 ]/H/I/C[0 1 1]}%
   user{/Subtype/Link/A<</Type/Action/S/URI/URI(#1)>>}%
  \relax
 }%
 \providecommand\@@endlink[0]{\pdfendlink}%
}%
\providecommand \url  [0]{\begingroup\@sanitize \@url }%
\providecommand \@url [1]{\endgroup\@href {#1}{\urlprefix}}%
\providecommand \urlprefix [0]{URL }%
\providecommand \Eprint[0]{\href }%
\@ifxundefined \urlstyle {%
  \providecommand \doi [1]{doi:\discretionary{}{}{}#1}%
}{%
  \providecommand \doi [0]{doi:\discretionary{}{}{}\begingroup
  \urlstyle{rm}\Url }%
}%
\providecommand \doibase [0]{http://dx.doi.org/}%
\providecommand \Doi[1]{\href{\doibase#1}}%
\providecommand \bibAnnote [3]{%
  \BibitemShut{#1}%
  \begin{quotation}\noindent
    \textsc{Key:}\ #2\\\textsc{Annotation:}\ #3%
  \end{quotation}%
}%
\providecommand \bibAnnoteFile [2]{%
  \IfFileExists{#2}{\bibAnnote {#1} {#2} {\input{#2}}}{}%
}%
\providecommand \typeout [0]{\immediate \write \m@ne }%
\providecommand \selectlanguage [0]{\@gobble}%
\providecommand \bibinfo [0]{\@secondoftwo}%
\providecommand \bibfield [0]{\@secondoftwo}%
\providecommand \translation [1]{[#1]}%
\providecommand \BibitemOpen[0]{}%
\providecommand \bibitemStop [0]{}%
\providecommand \bibitemNoStop [0]{.\EOS\space}%
\providecommand \EOS [0]{\spacefactor3000\relax}%
\providecommand \BibitemShut [1]{\csname bibitem#1\endcsname}%
\bibitem{yang:09}%
  \BibitemOpen
  \bibfield{author}{%
  \bibinfo {author} {\bibfnamefont{W.~L.}\ \bibnamefont{Yang}}, \bibinfo
  {author} {\bibfnamefont{A.~P.}\ \bibnamefont{Sorini}}, \bibinfo {author}
  {\bibfnamefont{C.-C.}\ \bibnamefont{Chen}}, \bibinfo {author}
  {\bibfnamefont{B.}~\bibnamefont{Moritz}}, \bibinfo {author}
  {\bibfnamefont{W.-S.}\ \bibnamefont{Lee}}, \bibinfo {author}
  {\bibfnamefont{F.}~\bibnamefont{Vernay}}, \bibinfo {author}
  {\bibfnamefont{P.}~\bibnamefont{Olalde-Velasco}}, \bibinfo {author}
  {\bibfnamefont{J.~D.}\ \bibnamefont{Denlinger}}, \bibinfo {author}
  {\bibfnamefont{B.}~\bibnamefont{Delley}}, \bibinfo {author}
  {\bibfnamefont{J.-H.}\ \bibnamefont{Chu}}, \bibinfo {author}
  {\bibfnamefont{J.~G.}\ \bibnamefont{Analytis}}, \bibinfo {author}
  {\bibfnamefont{I.~R.}\ \bibnamefont{Fisher}}, \bibinfo {author}
  {\bibfnamefont{Z.~A.}\ \bibnamefont{Ren}}, \bibinfo {author}
  {\bibfnamefont{J.}~\bibnamefont{Yang}}, \bibinfo {author}
  {\bibfnamefont{W.}~\bibnamefont{Lu}}, \bibinfo {author}
  {\bibfnamefont{Z.~X.}\ \bibnamefont{Zhao}}, \bibinfo {author}
  {\bibfnamefont{J.}~\bibnamefont{van~den Brink}}, \bibinfo {author}
  {\bibfnamefont{Z.}~\bibnamefont{Hussain}}, \bibinfo {author}
  {\bibfnamefont{Z.-X.}\ \bibnamefont{Shen}},\ and\ \bibinfo {author}
  {\bibfnamefont{T.~P.}\ \bibnamefont{Devereaux}},\ }%
  \bibfield{journal}{%
  \bibinfo {journal} {Phys. Rev. B}\ }%
  \textbf{\bibinfo {volume} {80}},\ \bibinfo {pages} {014508} (\bibinfo {year}
  {2009})%
  \bibAnnoteFile{NoStop}{yang:09}%
\bibitem{kaneko:08}%
  \BibitemOpen
  \bibfield{author}{%
  \bibinfo {author} {\bibfnamefont{K.}~\bibnamefont{Kaneko}}, \bibinfo {author}
  {\bibfnamefont{A.}~\bibnamefont{Hoser}}, \bibinfo {author}
  {\bibfnamefont{N.}~\bibnamefont{Caroca-Canales}}, \bibinfo {author}
  {\bibfnamefont{A.}~\bibnamefont{Jesche}}, \bibinfo {author}
  {\bibfnamefont{C.}~\bibnamefont{Krellner}}, \bibinfo {author}
  {\bibfnamefont{O.}~\bibnamefont{Stockert}},\ and\ \bibinfo {author}
  {\bibfnamefont{C.}~\bibnamefont{Geibel}},\ }%
  \bibfield{journal}{%
  \bibinfo {journal} {Phys. Rev. B}\ }%
  \textbf{\bibinfo {volume} {78}},\ \bibinfo {pages} {212502} (\bibinfo {year}
  {2008})%
  \bibAnnoteFile{NoStop}{kaneko:08}%
\bibitem{yildirim:08}%
  \BibitemOpen
  \bibfield{author}{%
  \bibinfo {author} {\bibfnamefont{T.}~\bibnamefont{Yildirim}},\ }%
  \bibfield{journal}{%
  \bibinfo {journal} {Phys. Rev. Lett.}\ }%
  \textbf{\bibinfo {volume} {101}},\ \bibinfo {pages} {057010} (\bibinfo {year}
  {2008})%
  \bibAnnoteFile{NoStop}{yildirim:08}%
\bibitem{yaresko:09}%
  \BibitemOpen
  \bibfield{author}{%
  \bibinfo {author} {\bibfnamefont{A.~N.}\ \bibnamefont{Yaresko}}, \bibinfo
  {author} {\bibfnamefont{G.-Q.}\ \bibnamefont{Liu}}, \bibinfo {author}
  {\bibfnamefont{V.~N.}\ \bibnamefont{Antonov}},\ and\ \bibinfo {author}
  {\bibfnamefont{O.~K.}\ \bibnamefont{Andersen}},\ }%
  \bibfield{journal}{%
  \bibinfo {journal} {Phys. Rev. B}\ }%
  \textbf{\bibinfo {volume} {79}},\ \bibinfo {pages} {144421} (\bibinfo {year}
  {2009})%
  \bibAnnoteFile{NoStop}{yaresko:09}%
\bibitem{han:09}%
  \BibitemOpen
  \bibfield{author}{%
  \bibinfo {author} {\bibfnamefont{M.~J.}\ \bibnamefont{Han}}, \bibinfo
  {author} {\bibfnamefont{Q.}~\bibnamefont{Yin}}, \bibinfo {author}
  {\bibfnamefont{W.~E.}\ \bibnamefont{Pickett}},\ and\ \bibinfo {author}
  {\bibfnamefont{S.~Y.}\ \bibnamefont{Savrasov}},\ }%
  \bibfield{journal}{%
  \bibinfo {journal} {Phys. Rev. Lett.}\ }%
  \textbf{\bibinfo {volume} {102}},\ \bibinfo {pages} {107003} (\bibinfo {year}
  {2009})%
  \bibAnnoteFile{NoStop}{han:09}%
\bibitem{ewings:08}%
  \BibitemOpen
  \bibfield{author}{%
  \bibinfo {author} {\bibfnamefont{R.~A.}\ \bibnamefont{Ewings}}, \bibinfo
  {author} {\bibfnamefont{T.~G.}\ \bibnamefont{Perring}}, \bibinfo {author}
  {\bibfnamefont{R.~I.}\ \bibnamefont{Bewley}}, \bibinfo {author}
  {\bibfnamefont{T.}~\bibnamefont{Guidi}}, \bibinfo {author}
  {\bibfnamefont{M.~J.}\ \bibnamefont{Pitcher}}, \bibinfo {author}
  {\bibfnamefont{D.~R.}\ \bibnamefont{Parker}}, \bibinfo {author}
  {\bibfnamefont{S.~J.}\ \bibnamefont{Clarke}},\ and\ \bibinfo {author}
  {\bibfnamefont{A.~T.}\ \bibnamefont{Boothroyd}},\ }%
  \bibfield{journal}{%
  \bibinfo {journal} {Phys. Rev. B}\ }%
  \textbf{\bibinfo {volume} {78}},\ \bibinfo {pages} {220501(R)} (\bibinfo
  {year} {2008})%
  \bibAnnoteFile{NoStop}{ewings:08}%
\bibitem{mcqueeney:08}%
  \BibitemOpen
  \bibfield{author}{%
  \bibinfo {author} {\bibfnamefont{R.~J.}\ \bibnamefont{McQueeney}}, \bibinfo
  {author} {\bibfnamefont{S.~O.}\ \bibnamefont{Diallo}}, \bibinfo {author}
  {\bibfnamefont{V.~P.}\ \bibnamefont{Antropov}}, \bibinfo {author}
  {\bibfnamefont{G.~D.}\ \bibnamefont{Samolyuk}}, \bibinfo {author}
  {\bibfnamefont{C.}~\bibnamefont{Broholm}}, \bibinfo {author}
  {\bibfnamefont{N.}~\bibnamefont{Ni}}, \bibinfo {author}
  {\bibfnamefont{S.}~\bibnamefont{Nandi}}, \bibinfo {author}
  {\bibfnamefont{M.}~\bibnamefont{Yethiraj}}, \bibinfo {author}
  {\bibfnamefont{J.~L.}\ \bibnamefont{Zarestky}}, \bibinfo {author}
  {\bibfnamefont{J.~J.}\ \bibnamefont{Pulikkotil}}, \bibinfo {author}
  {\bibfnamefont{A.}~\bibnamefont{Kreyssig}}, \bibinfo {author}
  {\bibfnamefont{M.~D.}\ \bibnamefont{Lumsden}}, \bibinfo {author}
  {\bibfnamefont{B.~N.}\ \bibnamefont{Harmon}}, \bibinfo {author}
  {\bibfnamefont{P.~C.}\ \bibnamefont{Canfield}},\ and\ \bibinfo {author}
  {\bibfnamefont{A.~I.}\ \bibnamefont{Goldman}},\ }%
  \bibfield{journal}{%
  \bibinfo {journal} {Phys. Rev. Lett.}\ }%
  \textbf{\bibinfo {volume} {101}},\ \bibinfo {pages} {227205} (\bibinfo {year}
  {2008})%
  \bibAnnoteFile{NoStop}{mcqueeney:08}%
\bibitem{zhao:08}%
  \BibitemOpen
  \bibfield{author}{%
  \bibinfo {author} {\bibfnamefont{J.}~\bibnamefont{Zhao}}, \bibinfo {author}
  {\bibfnamefont{D.-X.}\ \bibnamefont{Yao}}, \bibinfo {author}
  {\bibfnamefont{S.}~\bibnamefont{Li}}, \bibinfo {author}
  {\bibfnamefont{T.}~\bibnamefont{Hong}}, \bibinfo {author}
  {\bibfnamefont{Y.}~\bibnamefont{Chen}}, \bibinfo {author}
  {\bibfnamefont{S.}~\bibnamefont{Chang}}, \bibinfo {author}
  {\bibfnamefont{W.~R.}\ \bibnamefont{II}}, \bibinfo {author}
  {\bibfnamefont{J.~W.}\ \bibnamefont{Lynn}}, \bibinfo {author}
  {\bibfnamefont{H.~A.}\ \bibnamefont{Mook}}, \bibinfo {author}
  {\bibfnamefont{G.~F.}\ \bibnamefont{Chen}}, \bibinfo {author}
  {\bibfnamefont{J.~L.}\ \bibnamefont{Luo}}, \bibinfo {author}
  {\bibfnamefont{N.~L.}\ \bibnamefont{Wang}}, \bibinfo {author}
  {\bibfnamefont{E.~W.}\ \bibnamefont{Carlson}}, \bibinfo {author}
  {\bibfnamefont{J.}~\bibnamefont{Hu}},\ and\ \bibinfo {author}
  {\bibfnamefont{P.}~\bibnamefont{Dai}},\ }%
  \bibfield{journal}{%
  \bibinfo {journal} {Phys. Rev. Lett.}\ }%
  \textbf{\bibinfo {volume} {101}},\ \bibinfo {pages} {167203} (\bibinfo {year}
  {2008})%
  \bibAnnoteFile{NoStop}{zhao:08}%
\bibitem{diallo:09}%
  \BibitemOpen
  \bibfield{author}{%
  \bibinfo {author} {\bibfnamefont{S.~O.}\ \bibnamefont{Diallo}}, \bibinfo
  {author} {\bibfnamefont{V.~P.}\ \bibnamefont{Antropov}}, \bibinfo {author}
  {\bibfnamefont{T.~G.}\ \bibnamefont{Perring}}, \bibinfo {author}
  {\bibfnamefont{C.}~\bibnamefont{Broholm}}, \bibinfo {author}
  {\bibfnamefont{J.~J.}\ \bibnamefont{Pulikkotil}}, \bibinfo {author}
  {\bibfnamefont{N.}~\bibnamefont{Ni}}, \bibinfo {author}
  {\bibfnamefont{S.~L.}\ \bibnamefont{Bud'ko}}, \bibinfo {author}
  {\bibfnamefont{P.~C.}\ \bibnamefont{Canfield}}, \bibinfo {author}
  {\bibfnamefont{A.}~\bibnamefont{Kreyssig}}, \bibinfo {author}
  {\bibfnamefont{A.~I.}\ \bibnamefont{Goldman}},\ and\ \bibinfo {author}
  {\bibfnamefont{R.~J.}\ \bibnamefont{McQueeney}},\ }%
  \bibfield{journal}{%
  \bibinfo {journal} {Phys. Rev. Lett.}\ }%
  \textbf{\bibinfo {volume} {102}},\ \bibinfo {pages} {187206} (\bibinfo {year}
  {2009})%
  \bibAnnoteFile{NoStop}{diallo:09}%
\bibitem{zhao:09}%
  \BibitemOpen
  \bibfield{author}{%
  \bibinfo {author} {\bibfnamefont{J.}~\bibnamefont{Zhao}}, \bibinfo {author}
  {\bibfnamefont{D.~T.}\ \bibnamefont{Adroja}}, \bibinfo {author}
  {\bibfnamefont{D.-X.}\ \bibnamefont{Yao}}, \bibinfo {author}
  {\bibfnamefont{R.}~\bibnamefont{Bewley}}, \bibinfo {author}
  {\bibfnamefont{S.}~\bibnamefont{Li}}, \bibinfo {author}
  {\bibfnamefont{X.~F.}\ \bibnamefont{Wang}}, \bibinfo {author}
  {\bibfnamefont{G.}~\bibnamefont{Wu}}, \bibinfo {author}
  {\bibfnamefont{X.~H.}\ \bibnamefont{Chen}}, \bibinfo {author}
  {\bibfnamefont{J.}~\bibnamefont{Hu}},\ and\ \bibinfo {author}
  {\bibfnamefont{P.}~\bibnamefont{Dai}},\ }%
  \bibfield{journal}{%
  \bibinfo {journal} {Nat. Phys.}\ }%
  \textbf{\bibinfo {volume} {5}},\ \bibinfo {pages} {555} (\bibinfo {year}
  {2009})%
  \bibAnnoteFile{NoStop}{zhao:09}%
\bibitem{yin:08}%
  \BibitemOpen
  \bibfield{author}{%
  \bibinfo {author} {\bibfnamefont{Z.~P.}\ \bibnamefont{Yin}}, \bibinfo
  {author} {\bibfnamefont{S.}~\bibnamefont{Lebegue}}, \bibinfo {author}
  {\bibfnamefont{M.~J.}\ \bibnamefont{Han}}, \bibinfo {author}
  {\bibfnamefont{B.~P.}\ \bibnamefont{Neal}}, \bibinfo {author}
  {\bibfnamefont{S.~Y.}\ \bibnamefont{Savrasov}},\ and\ \bibinfo {author}
  {\bibfnamefont{W.~E.}\ \bibnamefont{Pickett}},\ }%
  \bibfield{journal}{%
  \bibinfo {journal} {Phys. Rev. Lett.}\ }%
  \textbf{\bibinfo {volume} {101}},\ \bibinfo {pages} {047001} (\bibinfo {year}
  {2008})%
  \bibAnnoteFile{NoStop}{yin:08}%
\bibitem{zhou:09}%
  \BibitemOpen
  \bibfield{author}{%
  \bibinfo {author} {\bibfnamefont{S.}~\bibnamefont{Zhou}}\ and\ \bibinfo
  {author} {\bibfnamefont{Z.}~\bibnamefont{Wang}},\ }%
  \bibinfo {note} {arXiv:0910.2707}%
  \bibAnnoteFile{NoStop}{zhou:09}%
\bibitem{stollhoff:81}%
  \BibitemOpen
  \bibfield{author}{%
  \bibinfo {author} {\bibfnamefont{G.}~\bibnamefont{Stollhoff}}\ and\ \bibinfo
  {author} {\bibfnamefont{P.}~\bibnamefont{Thalmeier}},\ }%
  \bibfield{journal}{%
  \bibinfo {journal} {Z. Phys. B}\ }%
  \textbf{\bibinfo {volume} {43}},\ \bibinfo {pages} {13} (\bibinfo {year}
  {1981})%
  \bibAnnoteFile{NoStop}{stollhoff:81}%
\bibitem{oles:81}%
  \BibitemOpen
  \bibfield{author}{%
  \bibinfo {author} {\bibfnamefont{A.~M.}\ \bibnamefont{Oles}},\ }%
  \bibfield{journal}{%
  \bibinfo {journal} {Phys. Rev. B}\ }%
  \textbf{\bibinfo {volume} {23}},\ \bibinfo {pages} {271} (\bibinfo {year}
  {1981})%
  \bibAnnoteFile{NoStop}{oles:81}%
\bibitem{collins:69}%
  \BibitemOpen
  \bibfield{author}{%
  \bibinfo {author} {\bibfnamefont{M.~F.}\ \bibnamefont{Collins}}, \bibinfo
  {author} {\bibfnamefont{V.~J.}\ \bibnamefont{Minkiewicz}}, \bibinfo {author}
  {\bibfnamefont{R.}~\bibnamefont{Nathans}}, \bibinfo {author}
  {\bibfnamefont{L.}~\bibnamefont{Passell}},\ and\ \bibinfo {author}
  {\bibfnamefont{G.}~\bibnamefont{Shirane}},\ }%
  \bibfield{journal}{%
  \bibinfo {journal} {Phys. Rev.}\ }%
  \textbf{\bibinfo {volume} {179}},\ \bibinfo {pages} {417} (\bibinfo {year}
  {1969})%
  \bibAnnoteFile{NoStop}{collins:69}%
\bibitem{zhai:09}%
  \BibitemOpen
  \bibfield{author}{%
  \bibinfo {author} {\bibfnamefont{H.}~\bibnamefont{Zhai}}, \bibinfo {author}
  {\bibfnamefont{F.}~\bibnamefont{Wang}},\ and\ \bibinfo {author}
  {\bibfnamefont{D.-H.}\ \bibnamefont{Lee}},\ }%
  \bibfield{journal}{%
  \bibinfo {journal} {Phys. Rev. B}\ }%
  \textbf{\bibinfo {volume} {80}},\ \bibinfo {pages} {064517} (\bibinfo {year}
  {2009})%
  \bibAnnoteFile{NoStop}{zhai:09}%
\bibitem{kou:09}%
  \BibitemOpen
  \bibfield{author}{%
  \bibinfo {author} {\bibfnamefont{S.-P.}\ \bibnamefont{Kou}}, \bibinfo
  {author} {\bibfnamefont{T.}~\bibnamefont{Li}},\ and\ \bibinfo {author}
  {\bibfnamefont{Z.-Y.}\ \bibnamefont{Weng}},\ }%
  \bibfield{journal}{%
  \bibinfo {journal} {Europhys. Lett.}\ }%
  \textbf{\bibinfo {volume} {88}},\ \bibinfo {pages} {17010} (\bibinfo {year}
  {2009})%
  \bibAnnoteFile{NoStop}{kou:09}%
\bibitem{medici:09}%
  \BibitemOpen
  \bibfield{author}{%
  \bibinfo {author} {\bibfnamefont{L.}~\bibnamefont{de'Medici}}, \bibinfo
  {author} {\bibfnamefont{S.}~\bibnamefont{Hassan}},\ and\ \bibinfo {author}
  {\bibfnamefont{M.}~\bibnamefont{Capone}},\ }%
  \bibfield{journal}{%
  \bibinfo {journal} {J. Supercond. Nov. Magn.}\ }%
  \textbf{\bibinfo {volume} {22}},\ \bibinfo {pages} {535} (\bibinfo {year}
  {2009})%
  \bibAnnoteFile{NoStop}{medici:09}%
\bibitem{krueger:09}%
  \BibitemOpen
  \bibfield{author}{%
  \bibinfo {author} {\bibfnamefont{F.}~\bibnamefont{Kr\"uger}}, \bibinfo
  {author} {\bibfnamefont{S.}~\bibnamefont{Kumar}}, \bibinfo {author}
  {\bibfnamefont{J.}~\bibnamefont{Zaanen}},\ and\ \bibinfo {author}
  {\bibfnamefont{J.}~\bibnamefont{van~den Brink}},\ }%
  \bibfield{journal}{%
  \bibinfo {journal} {Phys. Rev. B}\ }%
  \textbf{\bibinfo {volume} {79}},\ \bibinfo {pages} {054504} (\bibinfo {year}
  {2009})%
  \bibAnnoteFile{NoStop}{krueger:09}%
\bibitem{lv:10}%
  \BibitemOpen
  \bibfield{author}{%
  \bibinfo {author} {\bibfnamefont{W.}~\bibnamefont{Lv}}, \bibinfo {author}
  {\bibfnamefont{F.}~\bibnamefont{Kr{\"u}ger}},\ and\ \bibinfo {author}
  {\bibfnamefont{P.}~\bibnamefont{Phillips}},\ }%
  \bibinfo {note} {{arXiv}:1002.3165}%
  \bibAnnoteFile{NoStop}{lv:10}%
\bibitem{chen:09}%
  \BibitemOpen
  \bibfield{author}{%
  \bibinfo {author} {\bibfnamefont{C.-C.}\ \bibnamefont{Chen}}, \bibinfo
  {author} {\bibfnamefont{B.}~\bibnamefont{Moritz}}, \bibinfo {author}
  {\bibfnamefont{J.}~\bibnamefont{van~den Brink}}, \bibinfo {author}
  {\bibfnamefont{T.~P.}\ \bibnamefont{Devereaux}},\ and\ \bibinfo {author}
  {\bibfnamefont{R.~R.~P.}\ \bibnamefont{Singh}},\ }%
  \bibfield{journal}{%
  \bibinfo {journal} {Phys. Rev. B}\ }%
  \textbf{\bibinfo {volume} {80}},\ \bibinfo {pages} {180418(R)} (\bibinfo
  {year} {2009})%
  \bibAnnoteFile{NoStop}{chen:09}%
\bibitem{si:08}%
  \BibitemOpen
  \bibfield{author}{%
  \bibinfo {author} {\bibfnamefont{Q.}~\bibnamefont{Si}}\ and\ \bibinfo
  {author} {\bibfnamefont{E.}~\bibnamefont{Abrahams}},\ }%
  \bibfield{journal}{%
  \bibinfo {journal} {Phys. Rev. Lett.}\ }%
  \textbf{\bibinfo {volume} {101}},\ \bibinfo {pages} {076401} (\bibinfo {year}
  {2008})%
  \bibAnnoteFile{NoStop}{si:08}%
\bibitem{rodriguez:09}%
  \BibitemOpen
  \bibfield{author}{%
  \bibinfo {author} {\bibfnamefont{J.~P.}\ \bibnamefont{Rodriguez}}\ and\
  \bibinfo {author} {\bibfnamefont{E.~H.}\ \bibnamefont{Rezayi}},\ }%
  \bibfield{journal}{%
  \bibinfo {journal} {Phys. Rev. Lett.}\ }%
  \textbf{\bibinfo {volume} {103}},\ \bibinfo {pages} {097204} (\bibinfo {year}
  {2009})%
  \bibAnnoteFile{NoStop}{rodriguez:09}%
\bibitem{rodriguez:10}%
  \BibitemOpen
  \bibfield{author}{%
  \bibinfo {author} {\bibfnamefont{J.~P.}\ \bibnamefont{Rodriguez}},\ }%
  \bibinfo {note} {{arXiv}:1002.0891}%
  \bibAnnoteFile{NoStop}{rodriguez:10}%
\bibitem{yu:09}%
  \BibitemOpen
  \bibfield{author}{%
  \bibinfo {author} {\bibfnamefont{R.}~\bibnamefont{Yu}}, \bibinfo {author}
  {\bibfnamefont{K.~T.}\ \bibnamefont{Trinh}}, \bibinfo {author}
  {\bibfnamefont{A.}~\bibnamefont{Moreo}}, \bibinfo {author}
  {\bibfnamefont{M.}~\bibnamefont{Daghofer}}, \bibinfo {author}
  {\bibfnamefont{J.~A.}\ \bibnamefont{Riera}}, \bibinfo {author}
  {\bibfnamefont{S.}~\bibnamefont{Haas}},\ and\ \bibinfo {author}
  {\bibfnamefont{E.}~\bibnamefont{Dagotto}},\ }%
  \bibfield{journal}{%
  \bibinfo {journal} {Phys. Rev. B}\ }%
  \textbf{\bibinfo {volume} {79}},\ \bibinfo {pages} {104510} (\bibinfo {year}
  {2009})%
  \bibAnnoteFile{NoStop}{yu:09}%
\bibitem{eremin:09}%
  \BibitemOpen
  \bibfield{author}{%
  \bibinfo {author} {\bibfnamefont{I.}~\bibnamefont{Eremin}}\ and\ \bibinfo
  {author} {\bibfnamefont{A.~V.}\ \bibnamefont{Chubukov}},\ }%
  \bibfield{journal}{%
  \bibinfo {journal} {Phys. Rev. B}\ }%
  \textbf{\bibinfo {volume} {81}},\ \bibinfo {pages} {024511} (\bibinfo {year}
  {2010})%
  \bibAnnoteFile{NoStop}{eremin:09}%
\bibitem{lee:09a}%
  \BibitemOpen
  \bibfield{author}{%
  \bibinfo {author} {\bibfnamefont{H.}~\bibnamefont{Lee}}, \bibinfo {author}
  {\bibfnamefont{Y.-Z.}\ \bibnamefont{Zhang}}, \bibinfo {author}
  {\bibfnamefont{H.~O.}\ \bibnamefont{Jeschke}},\ and\ \bibinfo {author}
  {\bibfnamefont{R.}~\bibnamefont{Valenti}},\ }%
  \bibinfo {note} {{arXiv}:0912.4024}%
  \bibAnnoteFile{NoStop}{lee:09a}%
\bibitem{kubo:09}%
  \BibitemOpen
  \bibfield{author}{%
  \bibinfo {author} {\bibfnamefont{K.}~\bibnamefont{Kubo}}\ and\ \bibinfo
  {author} {\bibfnamefont{P.}~\bibnamefont{Thalmeier}},\ }%
  \bibfield{journal}{%
  \bibinfo {journal} {J. Phys. Soc. Jpn.}\ }%
  \textbf{\bibinfo {volume} {78}},\ \bibinfo {pages} {083704} (\bibinfo {year}
  {2009})%
  \bibAnnoteFile{NoStop}{kubo:09}%
\bibitem{daghofer:09}%
  \BibitemOpen
  \bibfield{author}{%
  \bibinfo {author} {\bibfnamefont{M.}~\bibnamefont{Daghofer}}, \bibinfo
  {author} {\bibfnamefont{A.}~\bibnamefont{Nicholson}}, \bibinfo {author}
  {\bibfnamefont{A.}~\bibnamefont{Moreo}},\ and\ \bibinfo {author}
  {\bibfnamefont{E.}~\bibnamefont{Dagotto}},\ }%
  \bibfield{journal}{%
  \bibinfo {journal} {Phys. Rev. B}\ }%
  \textbf{\bibinfo {volume} {81}},\ \bibinfo {pages} {014511} (\bibinfo {year}
  {2010})%
  \bibAnnoteFile{NoStop}{daghofer:09}%
\bibitem{uhrig:09}%
  \BibitemOpen
  \bibfield{author}{%
  \bibinfo {author} {\bibfnamefont{G.~S.}\ \bibnamefont{Uhrig}}, \bibinfo
  {author} {\bibfnamefont{M.}~\bibnamefont{Holt}}, \bibinfo {author}
  {\bibfnamefont{J.}~\bibnamefont{Oitmaa}}, \bibinfo {author}
  {\bibfnamefont{O.~P.}\ \bibnamefont{Sushkov}},\ and\ \bibinfo {author}
  {\bibfnamefont{R.~R.~P.}\ \bibnamefont{Singh}},\ }%
  \bibfield{journal}{%
  \bibinfo {journal} {Phys. Rev. B}\ }%
  \textbf{\bibinfo {volume} {79}},\ \bibinfo {pages} {092416} (\bibinfo {year}
  {2009})%
  \bibAnnoteFile{NoStop}{uhrig:09}%
\bibitem{yao:08}%
  \BibitemOpen
  \bibfield{author}{%
  \bibinfo {author} {\bibfnamefont{D.-X.}\ \bibnamefont{Yao}}\ and\ \bibinfo
  {author} {\bibfnamefont{E.~W.}\ \bibnamefont{Carlson}},\ }%
  \bibfield{journal}{%
  \bibinfo {journal} {Phys. Rev. B.}\ }%
  \textbf{\bibinfo {volume} {78}},\ \bibinfo {pages} {052507} (\bibinfo {year}
  {2008})%
  \bibAnnoteFile{NoStop}{yao:08}%
\bibitem{yao:09}%
  \BibitemOpen
  \bibfield{author}{%
  \bibinfo {author} {\bibfnamefont{D.-X.}\ \bibnamefont{Yao}}\ and\ \bibinfo
  {author} {\bibfnamefont{E.~W.}\ \bibnamefont{Carlson}},\ }%
  \bibinfo {note} {arXiv:0910.2528, special issue of Fron. Phys China}%
  \bibAnnoteFile{NoStop}{yao:09}%
\bibitem{smerald:09}%
  \BibitemOpen
  \bibfield{author}{%
  \bibinfo {author} {\bibfnamefont{A.}~\bibnamefont{Smerald}}\ and\ \bibinfo
  {author} {\bibfnamefont{N.}~\bibnamefont{Shannon}},\ }%
  \bibinfo {note} {{arXiv}:0909.2207}%
  \bibAnnoteFile{NoStop}{smerald:09}%
\bibitem{applegate:10}%
  \BibitemOpen
  \bibfield{author}{%
  \bibinfo {author} {\bibfnamefont{R.}~\bibnamefont{Applegate}}, \bibinfo
  {author} {\bibfnamefont{J.}~\bibnamefont{Oitmaa}},\ and\ \bibinfo {author}
  {\bibfnamefont{R.~R.~P.}\ \bibnamefont{Singh}},\ }%
  \bibfield{journal}{%
  \bibinfo {journal} {Phys. Rev. B}\ }%
  \textbf{\bibinfo {volume} {81}},\ \bibinfo {pages} {024505} (\bibinfo {year}
  {2010})%
  \bibAnnoteFile{NoStop}{applegate:10}%
\bibitem{schmidt:07b}%
  \BibitemOpen
  \bibfield{author}{%
  \bibinfo {author} {\bibfnamefont{B.}~\bibnamefont{Schmidt}}, \bibinfo
  {author} {\bibfnamefont{P.}~\bibnamefont{Thalmeier}},\ and\ \bibinfo {author}
  {\bibfnamefont{N.}~\bibnamefont{Shannon}},\ }%
  \bibfield{journal}{%
  \bibinfo {journal} {Phys. Rev. B}\ }%
  \textbf{\bibinfo {volume} {76}},\ \bibinfo {pages} {125113} (\bibinfo {year}
  {2007})%
  \bibAnnoteFile{NoStop}{schmidt:07b}%
\bibitem{thalmeier:08}%
  \BibitemOpen
  \bibfield{author}{%
  \bibinfo {author} {\bibfnamefont{P.}~\bibnamefont{Thalmeier}}, \bibinfo
  {author} {\bibfnamefont{M.~E.}\ \bibnamefont{Zhitomirsky}}, \bibinfo {author}
  {\bibfnamefont{B.}~\bibnamefont{Schmidt}},\ and\ \bibinfo {author}
  {\bibfnamefont{N.}~\bibnamefont{Shannon}},\ }%
  \bibfield{journal}{%
  \bibinfo {journal} {Phys. Rev. B}\ }%
  \textbf{\bibinfo {volume} {77}},\ \bibinfo {pages} {104441} (\bibinfo {year}
  {2008})%
  \bibAnnoteFile{NoStop}{thalmeier:08}%
\bibitem{shannon:04}%
  \BibitemOpen
  \bibfield{author}{%
  \bibinfo {author} {\bibfnamefont{N.}~\bibnamefont{Shannon}}, \bibinfo
  {author} {\bibfnamefont{B.}~\bibnamefont{Schmidt}}, \bibinfo {author}
  {\bibfnamefont{K.}~\bibnamefont{Penc}},\ and\ \bibinfo {author}
  {\bibfnamefont{P.}~\bibnamefont{Thalmeier}},\ }%
  \bibfield{journal}{%
  \bibinfo {journal} {Eur. Phys. J. B}\ }%
  \textbf{\bibinfo {volume} {38}},\ \bibinfo {pages} {599} (\bibinfo {year}
  {2004})%
  \bibAnnoteFile{NoStop}{shannon:04}%
\bibitem{klingeler:09}%
  \BibitemOpen
  \bibfield{author}{%
  \bibinfo {author} {\bibfnamefont{R.}~\bibnamefont{Klingeler}}, \bibinfo
  {author} {\bibfnamefont{N.}~\bibnamefont{Leps}}, \bibinfo {author}
  {\bibfnamefont{I.}~\bibnamefont{Hellmann}}, \bibinfo {author}
  {\bibfnamefont{A.}~\bibnamefont{Popa}}, \bibinfo {author}
  {\bibfnamefont{U.}~\bibnamefont{Stockert}}, \bibinfo {author}
  {\bibfnamefont{C.}~\bibnamefont{Hess}}, \bibinfo {author}
  {\bibfnamefont{V.}~\bibnamefont{Kataev}}, \bibinfo {author}
  {\bibfnamefont{H.-J.}\ \bibnamefont{Grafe}}, \bibinfo {author}
  {\bibfnamefont{F.}~\bibnamefont{Hammerath}}, \bibinfo {author}
  {\bibfnamefont{G.}~\bibnamefont{Lang}}, \bibinfo {author}
  {\bibfnamefont{S.}~\bibnamefont{Wurmehl}}, \bibinfo {author}
  {\bibfnamefont{G.}~\bibnamefont{Behr}}, \bibinfo {author}
  {\bibfnamefont{L.}~\bibnamefont{Harnagea}}, \bibinfo {author}
  {\bibfnamefont{S.}~\bibnamefont{Singh}},\ and\ \bibinfo {author}
  {\bibfnamefont{B.}~\bibnamefont{B\"uchner}},\ }%
  \bibfield{journal}{%
  \bibinfo {journal} {Phys. Rev. B}\ }%
  \textbf{\bibinfo {volume} {81}},\ \bibinfo {pages} {024506} (\bibinfo {year}
  {2010})%
  \bibAnnoteFile{NoStop}{klingeler:09}%
\bibitem{sales:09}%
  \BibitemOpen
  \bibfield{author}{%
  \bibinfo {author} {\bibfnamefont{B.}~\bibnamefont{Sales}}, \bibinfo {author}
  {\bibfnamefont{M.}~\bibnamefont{McGuire}}, \bibinfo {author}
  {\bibfnamefont{A.}~\bibnamefont{Sefat}},\ and\ \bibinfo {author}
  {\bibfnamefont{D.}~\bibnamefont{Mandrus}},\ }%
  \bibfield{journal}{%
  \bibinfo {journal} {Physica C: Superconductivity}\ }%
  \textbf{\bibinfo {volume} {470}},\ \bibinfo {pages} {304 } (\bibinfo {year}
  {2010})%
  \bibAnnoteFile{NoStop}{sales:09}%
\bibitem{korshunov:09}%
  \BibitemOpen
  \bibfield{author}{%
  \bibinfo {author} {\bibfnamefont{M.~M.}\ \bibnamefont{Korshunov}}, \bibinfo
  {author} {\bibfnamefont{I.}~\bibnamefont{Eremin}}, \bibinfo {author}
  {\bibfnamefont{D.~V.}\ \bibnamefont{Efremov}}, \bibinfo {author}
  {\bibfnamefont{D.~L.}\ \bibnamefont{Maslov}},\ and\ \bibinfo {author}
  {\bibfnamefont{A.~V.}\ \bibnamefont{Chubukov}},\ }%
  \bibfield{journal}{%
  \bibinfo {journal} {Phys. Rev. Lett.}\ }%
  \textbf{\bibinfo {volume} {102}},\ \bibinfo {pages} {236403} (\bibinfo {year}
  {2009})%
  \bibAnnoteFile{NoStop}{korshunov:09}%
\bibitem{cricchio:09}%
  \BibitemOpen
  \bibfield{author}{%
  \bibinfo {author} {\bibfnamefont{F.}~\bibnamefont{Cricchio}}, \bibinfo
  {author} {\bibfnamefont{O.}~\bibnamefont{Gran{\"a}s}},\ and\ \bibinfo
  {author} {\bibfnamefont{L.}~\bibnamefont{Nordstr{\"o}m}},\ }%
  \bibinfo {note} {arXiv:0911.1342}%
  \bibAnnoteFile{NoStop}{cricchio:09}%
\bibitem{lee:09}%
  \BibitemOpen
  \bibfield{author}{%
  \bibinfo {author} {\bibfnamefont{C.-C.}\ \bibnamefont{Lee}}, \bibinfo
  {author} {\bibfnamefont{W.-G.}\ \bibnamefont{Yin}},\ and\ \bibinfo {author}
  {\bibfnamefont{W.}~\bibnamefont{Ku}},\ }%
  \bibfield{journal}{%
  \bibinfo {journal} {Phys. Rev. Lett.}\ }%
  \textbf{\bibinfo {volume} {103}},\ \bibinfo {pages} {267001} (\bibinfo {year}
  {2009})%
  \bibAnnoteFile{NoStop}{lee:09}%
\bibitem{zhang:10}%
  \BibitemOpen
  \bibfield{author}{%
  \bibinfo {author} {\bibfnamefont{Y.-Z.}\ \bibnamefont{Zhang}}, \bibinfo
  {author} {\bibfnamefont{I.}~\bibnamefont{Opahle}}, \bibinfo {author}
  {\bibfnamefont{H.~O.}\ \bibnamefont{Jeschke}},\ and\ \bibinfo {author}
  {\bibfnamefont{R.}~\bibnamefont{Valent\'\i{}}},\ }%
  \bibfield{journal}{%
  \bibinfo {journal} {Phys. Rev. B}\ }%
  \textbf{\bibinfo {volume} {81}},\ \bibinfo {pages} {094505} (\bibinfo {year}
  {2010})%
  \bibAnnoteFile{NoStop}{zhang:10}%
\bibitem{shannon:06}%
  \BibitemOpen
  \bibfield{author}{%
  \bibinfo {author} {\bibfnamefont{N.}~\bibnamefont{Shannon}}, \bibinfo
  {author} {\bibfnamefont{T.}~\bibnamefont{Momoi}},\ and\ \bibinfo {author}
  {\bibfnamefont{P.}~\bibnamefont{Sindzingre}},\ }%
  \bibfield{journal}{%
  \bibinfo {journal} {Phys. Rev. Lett.}\ }%
  \textbf{\bibinfo {volume} {96}},\ \bibinfo {pages} {027213} (\bibinfo {year}
  {2006})%
  \bibAnnoteFile{NoStop}{shannon:06}%
\bibitem{ohyama:94}%
  \BibitemOpen
  \bibfield{author}{%
  \bibinfo {author} {\bibfnamefont{T.}~\bibnamefont{Ohyama}}\ and\ \bibinfo
  {author} {\bibfnamefont{H.}~\bibnamefont{Shiba}},\ }%
  \bibfield{journal}{%
  \bibinfo {journal} {J. Phys. Soc. Jpn.}\ }%
  \textbf{\bibinfo {volume} {63}},\ \bibinfo {pages} {3454} (\bibinfo {year}
  {1994})%
  \bibAnnoteFile{NoStop}{ohyama:94}%
\bibitem{veillette:05}%
  \BibitemOpen
  \bibfield{author}{%
  \bibinfo {author} {\bibfnamefont{M.~Y.}\ \bibnamefont{Veillette}}, \bibinfo
  {author} {\bibfnamefont{J.~T.}\ \bibnamefont{Chalker}},\ and\ \bibinfo
  {author} {\bibfnamefont{R.}~\bibnamefont{Coldea}},\ }%
  \bibfield{journal}{%
  \bibinfo {journal} {Phys. Rev. B}\ }%
  \textbf{\bibinfo {volume} {71}},\ \bibinfo {pages} {214426} (\bibinfo {year}
  {2005})%
  \bibAnnoteFile{NoStop}{veillette:05}%
\end{thebibliography}%

\end{document}